\documentclass[twocolumn]{aastex631}
\usepackage{amsmath}
\usepackage{multirow}
\usepackage{amssymb}
\usepackage{graphicx}
\usepackage{color}
\usepackage{float}
\usepackage{comment}

\definecolor{darkgreen}{rgb}{0,0.35,0}
\usepackage{CJK}

\begin{document}
\begin{CJK*}{UTF8}{gbsn}

\title{Accretion of AGN Stars under Influence of Disk Geometry}

\shorttitle{3D RHD Simulations of AGN-Embedded Stars}
\shortauthors{Chen et al.}

\author[0000-0003-3792-2888]{Yi-Xian Chen (陈逸贤)}
\affiliation{Department of Astrophysical Sciences, Princeton University,  4 Ivy Lane, Princeton, NJ 08544, USA}

\author{Yan-Fei Jiang (姜燕飞)}
\affiliation{Center for Computational Astrophysics, Flatiron Institute, New York, NY 10010, USA}

\author{Jeremy Goodman}
\affiliation{Department of Astrophysical Sciences, Princeton University,  4 Ivy Lane, Princeton, NJ 08544, USA}

\begin{abstract}
Massive stars can form within or be captured by AGN disks, influencing both the thermal structure and metallicity of the disk environment. 
In a previous work, we investigated isotropic accretion onto massive stars from a gas-rich, high-entropy background. 
Here, 
we consider a more realistic scenario by incorporating the stratified geometry of the background disk in our 3D radiation hydrodynamic simulatons. 
We find that accretion remains relatively isotropic when the disk is hot enough and the scale height is thicker than the accretion flow's nominal supersonic critical radius $R_{\rm crit}$ (sub-thermal). 
However, when the disk becomes cold, the accretion flow becomes significantly anisotropic (super-thermal).
Escaping stellar and accretion luminosity can drive super-Eddington outflows in the polar region, 
while rapid accretion is sustained along the midplane. 
Eventually, 
the effective cross-section is constrained by the Hill radius and the disk scale height rather than the critical radius when the disk is cold enough. 
For our setup (stellar mass $\sim 50 M_\odot$ and background density $\rho\sim 10^{-10}$ g/cm$^3$) the accretion rates is capped below $\sim 0.02M_\odot$/year and the effective accretion parameter $\alpha\sim 10^{-1}$ over disk temperature range $3 - 7 \times 10^4$ K. Spiral arms facilitate inward mass flux by driving outward angular momentum transport. 
Gap-opening effects may further reduce the long-term accretion rate, 
albeit to confirm which requires global simulations evolved over much longer viscous timescales. 
\end{abstract}

\section{Introduction}
\label{sec:intro}

Active Galactic Nuclei (AGN) are powered by accretion disks around supermassive Black Holes (SMBHs) 
\citep{Lyndenbell1969}.
In the outskirts of AGN disks, 
gravitational instability can lead to
intense star formation from disk fragmentation \citep{Paczynski1978,Goodman2003,Levin2003,Jiang11,Chen2023}, likely even in the presence of strong magnetic fields \citep{guo2025idealizedglobalmodelsaccretion}. 
Moreover, 
stars originated from the nuclear cluster \citep{KormendyHo2013} can also be captured into the disk through gas drag \citep{Artymowicz1993,MacLeod2020,WangYH2024}. 
Stars evolving within the gas-rich midplane of AGN disks can grow to be more massive and luminous than typical stars evolving in isolation. 
These stars may act as progenitors of merging black holes in AGN disks, which could contribute to high progenitor mass merger events observed by LIGO-Virgo \citep{McKernan2012,McKernan2014,Tagawa2020a,Li2021,Samsing2022,Chen2022, Epstein-Martin2024}. Meanwhile, the chemical yields
of these stars can contribute to the super-solar metallicities observed in high-redshift quasars, 
as inferred from 
broad emission lines \citep{Hamann1999,Hamann2002,Nagao2006,Xu2018,Wang+2022,Lai+2022,Huang2023,Floris2024, Fryer2025}.

Recent studies \citep{Cantiello2021,Dittmann2021,AliDib2023} used 1D simulation code MESA \citep{MESA2011,MESA2013,MESA2018,MESA2019} to investigate stellar evolution in AGN disks over nuclear timescales, 
incorporating simplified models for stellar accretion from the background environment, as well as mass loss by trans-Eddington stellar wind. 
These works suggest that, 
in environments with sufficiently high ambient densities, 
stars can grow to equilibrium masses of
several hundred $M_\odot$ 
by balancing accretion with mass loss,
with the luminosity factor $\lambda_\star := L_\star/L_{\rm Edd} \approx 1$.

The accretion prescription currently applied in 1D models \citep{Cantiello2021,Dittmann2021,AliDib2023} is based on the Bondi rate calculated from characteristic AGN disk density and temperature parameters. 
In addition to this, 
the prescription takes into account radiative feedback from the stellar intrinsic luminosity, 
assuming that effective gravity is reduced by a fraction of $\lambda_\star$.
In a previous paper \citep{Chen2024}, 
we conducted radiation hydrodynamic simulations of accreting stellar envelopes in an isotropic background environment, mimicking the case where star's accretion radius is much smaller than the disk scale height, 
to conclude that 
i) In the fast-diffusion regime where radiation force effectively counteracts gravity, as assumed by the existing 1D prescription, 
additional feedback from radiation entropy release and gravitational energy dissipation within the accretion flow can \textit{further} reduce the accretion rate by more than one order of magnitude, 
with the latter scenario resembling Eddington-limited accretion of black holes $(\lambda_\star = 0)$.
ii) in the very optically thick regime when diffusion timescale is long ($c/\tau < c_s$, 
where $\tau$ is optical depth within 
the sonic radius for infall), 
radiation becomes coupled with gas, 
no longer acting as a simple reduction to stellar gravity and the accretion process becomes adiabatic.


In this work, 
we improve on the preceding simulations and consider outer boundary condition more realistic for AGN disks, namely when the stellar sphere of influence becomes comparable to the disk scale height. 
We account for the 
vertical stratification of density/temperature, 
the shear velocity of background Keplerian flow, 
as well as source terms for the tidal potential induced by the host SMBH and rotation of the frame. 
We focus on the fast-diffusion regime and find that the feedback factor can be strongly altitude-dependent. 
The feedback is low in the midplane where it's most optically thick and allows for rapid accretion, 
while diffusive luminosity can be super-Eddington in the polar region, driving strong outflows. This anisotropy is more pronounced when the background disk is cold and thin, such that the disk scale height is significantly smaller than the star's sphere of influence.

The paper is organized as follows:
In \S \ref{sec:setup}, 
we introduce our numerical setup, focusing on how our boundary condition is different from \citet{Chen2024}.
In \S \ref{sec:results}, 
we present the results obtained from our simulations.
We emphasize 
how the flow structure and measured accretion rates can deviate from what's expected in an isotropic setup due to influence of vertically stratified geometry.
We discuss some caveats of our short-term simulation in \S \ref{sec:discussions} and 
summarize our findings in \S \ref{sec:summary}. 
Subsequent works will cover the details of adiabatic accretion in the slow diffusion regime, 
and especially cases where the star could acquire a self-gravitating envelope.

\section{3D Numerical Setup}
\label{sec:setup}

We perform radiation hydrodynamic (RHD) simulations using the standard Godunov method in \texttt{Athena++} \citep{Stone2020}. 
In a spherical polar coordinate system $(r, \theta, \phi)$ centered on the accreting massive star, 
we
solve ideal hydrodynamic equations coupled with the time-dependent 
frequency-integrated 
implicit radiation transport equation for specific intensities over discrete angles \citep{Jiang2014,Jiang2021}. {This method efficiently and accurately handles the optically thick midplane, the optically thin disk surface, and the transition region in between \citep[e.g.,][]{Chen2023}.}

The basic physical parameters 
chosen are the same as in \citet{Chen2024}. 
The gas mean molecular weight is $\mu = 0.60 m_p$, 
assuming a fully ionized composition of $X = 0.73 , Y= 0.25, Z = 0.02$. 
The gas adiabatic index is $\gamma=5/3$, 
also consistent with the outer regions 
of MESA models developed in \citet{Cantiello2021,AliDib2023}. 
The opacities used in our simulation are interpolated from 
the OPAL tables \citep{Iglesias1996} 
for the same composition. 

In \citet{Chen2024}, the
gravity field is taken to be spherically symmetric with only contribution from the stellar core $\Phi_\star = -GM_\star/r$ \citep{Schultz+2022}, 
assuming that self-gravity of the outer envelope in the simulation domain is negligible. 
In this work, 
we further add the gravitational potential from the SMBH companion: 

\begin{equation}
    \Phi_\bullet = -\dfrac{GM_\bullet}{|{\bf r_\bullet} - { \bf r}|},
\end{equation}

Where ${\bf r_\bullet}  =  (r_\bullet, 0, \pi/2)$ is the location of the SMBH \footnote{In other words, the SMBH is placed along the $y$ axis for a corresponding right-handed Cartesian coordinate system. }. 
At an orbital distance of $r_\bullet$ around a SMBH of mass $M_\bullet \ll M_\star$, the disk angular velocity $\Omega_0 = \sqrt{GM_\bullet/r_\bullet^3}$ 
vector is pointed towards the vertical direction. The Hill radius of the companion can be calculated as  $R_{\rm Hill} = r_\bullet (M_\star/M_\bullet)^{1/3}$
To aid our analysis, 
we define a cylindrical coordinate system $(R, z, \varphi)$ based on $(r, \theta, \phi)$. 
To fix the location of the host, our simulation is carried out in a frame co-rotating 
at angular velocity $\Omega_0 \hat{z}$ with respect to the origin.
For this setup, we need to add the indirect and centrifugal terms

\begin{equation}
      \Phi_{\rm ind} = \dfrac{GM_\bullet}{r_\bullet^2} ({\bf r_\bullet}\cdot {\bf r}), \Phi_{\rm cen} = - \dfrac{1}{2} \Omega_0^2 R^2
\end{equation}

to the potential. The former effect is because the coordinate origin is $r_\bullet$ from the true center of mass of the binary system. 
We implement Coriolis force as described in the Appendix of \citet{Zhu2021}, who applies a similar setup as in our simulation to study the envelope of a proto-Jupiter embedded in a protostellar disk.

\subsection{Initial and Boundary Conditions}

We use the same initial conditions as in \citep{Chen2024}, 
a stellar profile based on a $50\mathrm{M_\odot}$, 
1D MESA stellar model with a luminosity of 
$8.6\times 10^5 L_\odot$ and photosphere radius of $\sim 40R_\odot$ by the end of its main-sequence. 
We set the lower boundary of our computational domain at $R_{\rm in} = 24.65 R_\odot$ at its outermost radiative zone, so 99\% of stellar mass is contained below $R_{\rm in}$ and can be modeled simply as a source term outside the simulation domain, 
while self-gravity of the envelope within the simulation domain can be neglected.
In our initial conditions, 
we modify the profile of the convective regions so as to support the stellar luminosity purely radiatively. 
This setup automatically results in convective instability in 3D. 
Once the simulation starts, the stellar profile self-consistently relaxes to a convective region more extended than predicted by MESA, 
up to a radius of $\sim 100R_\odot$. This expansion is typical in 3D simulations of convective stellar envelopes due to turbulent pressure support \citep{JiangReview2023}.
The radiative zone indicated by MESA between $\sim 30R_\odot$ and $R_{\rm in}$, 
however, would remain radiative and the velocity fluctuation there is significantly lower than in the convective zone. 
All initial density and temperature beyond the stellar photosphere are initially set to floor values.


Our simulation domain cover the full azimuthal range of $\phi \in [0, 2\pi]$, a polar range of $\theta \in [0, \pi/2]$ and a logarithmic radial grid that ranges $r\in [R_{\rm in} = 24.65 R_\odot, R_{\rm out}]$. 
{This ``shearing-globe” setup aims to extend previous stellar accretion studies in isotropic environments to a disk background, 
enabling controlled comparisons by varying only the external conditions—similar to the 1D and 3D comparison in \citet{Zhu2021}. 
An advantage is that it resolves the stellar convective envelope and its interaction with the accretion flow 
using a logarithmic radial grid, 
without requiring excessive mesh refinement near the inner boundary, as would be required for a more conventional Cartesian shearing-box setup.}

The density and temperature are fixed and the velocity is set to zero 
at the inner boundary in $r$ inside the stellar radiative zone. 
The outer boundary condition in $r$ is set by the following procedure: 
For a chosen value of $T_{0}$ and $\rho_{0}$ representing the \textit{midplane} density and temperature of the AGN disk background, 
the \textit{boundary} ghost cell profiles are fixed to

\begin{equation}
    \rho(z) = \rho_0 (1 - z^2\Omega^2 /8c_{s, 0}^2)^{3}, T(z) = T_0 (1 - z^2 \Omega^2/8c_{s, 0}^2),
    \label{eqn:verticalprofile}
\end{equation}

for $z < 2\sqrt{2} H := 2\sqrt{2}c_{s, 0}/\Omega$ or floor values at larger $z$. This is the solution 
of hydrostatic equilibrium 
neglecting $R$ dependence and considering only SMBH and rotational potential terms first order in $z^2$, 
while
assuming the background disk follows a polytropic with $P\propto \rho^{4/3}$ or constant radiation to gas pressure ratio $\Pi_0 = a_{\rm rad} T_{0}^3\mathcal{R}/\mu \rho_{0}$, 
similar to \citet{Chen2023} 
but without disk self-gravity. {This is also a nearly isentropic profile since the effective adiabatic index $\gamma \approx 4/3$ for a mixture of radiation and gas with $\Pi_0\gtrsim 1$, applicable to our choice of parameters.}
$c_{s, 0}$ is the characteristic midplane sound speed set by the combined pressure of radiation and gas, satisfying\footnote{Note that Equation \eqref{eq:cs0def} defines $c_{s,0}^2$ as $P/\rho$ rather than $(\partial P/\partial\rho)_S$}

\begin{equation}\label{eq:cs0def}
\begin{aligned}
    \rho_0 c_{s, 0}^2 &= \rho_0 c_{s,gas, 0}^2 + \rho_0 c_{s,rad, 0}^2 \\&= \dfrac{\rho_0 \mathcal{R} T_0}{\mu} + \dfrac{a_{\rm rad} T_0^4}{3} =  \dfrac{\rho_0 \mathcal{R} T_0}{\mu} (1+\Pi_0),
    \end{aligned}
\end{equation}

while $H$ is the scale height defined by this total sound speed.
For this vertical profile, 
the integrated surface density from midplane to the photosphere is $\Sigma \approx 2.5 \rho_{0} H$. We also scale the simulation outer boundary as $R_{\rm out}\sim 3H$ for different midplane temperature, 
to keep the initial free-fall timescale $R_{\rm out}/c_{s, 0}$ roughly constant.

The outer boundary velocity is set to be the background Keplerian velocity with respect to SMBH, 
while subtracting off the frame rotation:

\begin{equation}
    v = (\sqrt{GM_\bullet/|{\bf r_\bullet} - {\bf r}|} - \Omega_0 |{\bf r_\bullet} - {\bf r}|),
\end{equation}

which is pointed towards the direction normal to ${\bf r_\bullet} - {\bf r}$. 
The initial radiation flux is set to be isotropic consistent with a radiation energy density of $aT^4$. 
Although it may not be realistic to impose these constraints on velocity and radiation field above the photosphere  $z> 2\sqrt{2} H$ where it's very optically thin, 
the density there is also low enough such that these boundary conditions does not affect the accretion process. 
Also, 
below the photosphere, 
both the ``background" density and velocity profiles are not entirely in steady state when radial pressure gradient along ${\bf r_\bullet}$ is considered. 
However, 
as the disk gas is gradually ``injected" into the simulation domain via the shear velocity and fills up the midplane, 
the disk will eventually reach distances close to the stellar surface 
and further adjust under the influence of stellar gravity, 
reaching a new equilibrium distribution. 
Consequently, minor deviations in the initial vertical distribution have limited impact on the qualitative geometry of the accretion flow, 
which we discuss in \S \ref{sec:results}. {We resolve the domain with $N_\phi = 200, N_\theta = 50$ 
and an adjustable $N_r$ chosen to ensure  $\Delta r/r \approx \Delta \phi = \Delta \theta = 0.03$ up to $R_{\rm out}$ (e.g. $N_r = 120$ up to $R_{\rm out} = 855.6 R_\odot$ for $T_0 = 5\times 10^4$K). 
We also add one level of mesh refinement within 300 $R_\odot$.}


\subsection{Diagnostics}
\label{sec:diagnostics}

To facilitate analysis of our simulation results, 
we introduce notation for important average variables. 
In a quasi-steady state, 
we define the time and azimuthally-averaged \textit{2D} profiles as 

\begin{equation}
    \langle X  \rangle (r, \theta)= \dfrac{\int X (r, \theta, \phi, t) {\rm d}\phi {\rm d} t}{\int {\rm d}\phi {\rm d} t}
    \label{eq:azimuthal_averaging}
\end{equation}

In our analysis, we perform average over the last disk dynamical timescale $\Omega^{-1}$ of the simulations after they reach steady states.

To further obtain average radial profiles, we define

\begin{equation}
    \langle X  \rangle (r)= \dfrac{\int \langle X  \rangle (r, \theta) \sin \theta {\rm d}\theta }{\int \sin \theta  {\rm d}  \theta}
    \label{eq:averaging}
\end{equation}

We define the density-weighted square of the isothermal gas sound speed as 

\begin{equation}\label{eq:csdef}
    \langle c_{\rm s, gas}^2 \rangle_\rho = \dfrac{\langle P_{\rm gas} \rangle}{\langle \rho \rangle}
\end{equation}

While the radiation sound speed is 

\begin{equation}
    \langle c_{\rm s, rad}^2 \rangle_\rho = \dfrac{\langle P_{\rm rad}  \rangle}{\langle \rho \rangle}
\label{eq:csrad}
\end{equation}

When determining the sonic radius, it's important to compare the gas sound speed profiles with the radial velocity:

\begin{equation}
    \langle v_{r} \rangle_\rho = \dfrac{\langle \rho v_{r}  \rangle}{\langle \rho \rangle}
\end{equation}

Which is related to the average radial mass flux

\begin{equation}
   \langle \dot{M} \rangle= 4\pi r^2 {\langle \rho v_{r}  \rangle} = 4\pi r^2 \langle \rho\rangle \langle v_{r}  \rangle_\rho
\end{equation}

The radial energy flux of the accretion flow is contributed by radiation, 
gravitational potential, 
and gas energy. 
The radiative luminosity can be further decomposed into a diffusive term $ L_{\rm diff} = 4\pi r^2 \langle F_{\rm diff}  \cdot \hat{r} \rangle $ and a term representing the advection of radiative enthalpy $ L_{\rm adv} = 4\pi r^2 \langle F_{\rm adv} \cdot \hat{r} \rangle$. 
{To be more specific, 
$F_{\rm diff}$ and $F_{\rm adv}$ are inherently 3D vectors, defined as radiation fluxes in the co-moving frame and its difference to that measured the lab frame \citep{Jiang2021}:}

\begin{equation}
    F_{\rm diff} = \int I_{\rm com} \mathbf{n} d\Omega,  F_{\rm adv} = \int I_{\rm lab} \mathbf{n} d\Omega - \int I_{\rm com} \mathbf{n} d\Omega,
\end{equation}

{such that $F_{\rm diff}$ is the component exerting force on the comoving gas, 
while $F_{\rm adv}$ is contributed by advection of radiation with respect to the comoving frame. 
Here $I_{\rm com}$ and $ I_{\rm lab}$ are the comoving and lab frame intensities defined over solid angles with orientation $\mathbf{n}$. 
For optically thick regions, $F_{\rm adv}  = 4 P_{\rm rad} v$ }.

The ``luminosity" 
in the form of advected gravitational potential energy is 

\begin{equation}
    L_{\rm grav} =  \dfrac{GM_\star}{r} \langle \dot{M} \rangle(r)
\label{eq:lgrav},
\end{equation}

which scales with $1/r$ when $\langle \dot{M} \rangle (r)$ converges to a constant $\dot{M}$ in a quasi-steady state. Finally, the luminosity carried by gas thermal and kinetic  energy is 

\begin{equation}
    L_{\rm gas}  = 4\pi r^2  \langle (U_{\rm gas} + \rho v^2 ) v_r \rangle(r)
\end{equation}

This term usually plays a minor role in the energy budget, being at most comparable to a small fraction of the gravitational potential term.

Similar to \citet{Chen2024}, 
we define the sum of scattering and absorption opacity as $\kappa = \kappa_{\rm s} + \kappa_{\rm a}$.
We define the dimensionless measure of diffusive luminosity, 
characterizing the effect of reduced gravity

\begin{equation}
    \lambda_{\rm diff} =  
    \dfrac{r^2}{GM_\star c}  \langle {\kappa F_{\rm diff} \cdot \hat{r}} \rangle(r,\theta)
\end{equation}

Here we've introduced the Eddington luminosity $L_{\rm Edd} = 4\pi GM_\star c/\kappa $
and Eddington ratio $\lambda_{\rm diff}$ based on the local opacity $\kappa$. 
In the fast-diffusion regime, 
$\lambda_{\rm diff}$ indicates the fraction of radial gravity being canceled out by outward radiative force. 
As we will show below, 
$\lambda_{\rm diff} (r,\theta)$ can be very anisotopic when vertically stratified background structures are considered.

Another major difference our new setup compared to isotropic simulations of \citet{Chen2024} is 
the existence of a centrifugally supported circumstellar flow, through which most of the mass accretion occurs.
The net accretion of angular momentum through the disk must nearly vanish, else the disk could not persist.
So inward advective transport of angular momentum must be balanced by outward transport via some other mechanism.
In principle, viscosity or magnetorotational turbulence could do this, but neither was included in these simulations.
The only remaining possibility is Reynolds stress.
The question then becomes, what is the origin of this stress?
We will address this question in \S\ref{sec:AMtransport}, but here we merely describe how the advective and Reynolds angular-momentum fluxes are measured.

Across the surface of a sphere with radius $r$, 
the advective vertical angular momentum flux is 

\begin{equation}
    \dot{J}_{\rm adv}  =  
    4\pi r^2 \langle r  \overline{\rho v}_r  \overline{ v}_\phi  \sin\theta \rangle (r)
\end{equation}

{Note that for each snapshot before time and spatial average, 
we first decompose the radial mass flux $\rho v_r$ and $v_\phi$ as the sum of azimuthally averaged component $ \overline{\rho v_r} , \overline{v_\phi}$ and the deviation $ \delta{(\rho v_r)}, \delta{v_\phi}$. 
Since $\overline{\rho v}_r<0$ for mass accretion, $ \dot{J}_{\rm adv} <0$ also transports angular momentum inwards.
The angular momentum flux due to Reynolds stress is given by}

\begin{equation}
 \dot{J}_{\rm Rey} =  
  4\pi r^2  r R_{r\phi}(r) \sin\theta
  \label{eqn:Jvis}
\end{equation}

where $R_{ r\phi}$(r) is the Reynolds stress $\langle\delta (\rho v_{r}) \delta v_{\phi} \rangle$. This stress term gives rise to an effective ``standard disk" accretion parameter \citep{SS1973} of 

\begin{equation}
    \alpha_{\rm eff} = \dfrac{ \langle R_{r\phi} \sin\theta \rangle (r)}{\langle P_{\rm rad} + P_{\rm gas}\rangle (r)}
\end{equation}

In a steady-state accretion disk around the AGN star approaching axisymmetry, 
we can define the \textit{specific} angular momentum flux 
\begin{equation}
    \dot{\mathcal{J}}_{\rm adv} =  \dot{J}_{\rm adv}/\langle \dot{M} \rangle (r), \dot{\mathcal{J}}_{\rm Rey} = \dot{J}_{\rm Rey}/\langle \dot{M} \rangle (r),
\end{equation}
and we expect angular momentum transport to satisfy $\dot{\mathcal{J}}_{\rm Rey} + \dot{\mathcal{J}}_{\rm adv}  = {\rm Constant}$ within the Hill sphere. 
Full angular momentum conservation at a larger radial scale, 
however, requires considering differential gravitational torques, which become important approaching the truncation radius of the disk at $\sim R_H$. 
Notably, 
the assumption $\delta v \ll \overline{v}$ breaks down as the flow becomes dominated by the non-axisymmetric background shear and our calculation above becomes less meaningful. 
However, for analyzing the rotational aspect of the circum-stellar flow within $R_H$, 
measurement of $\dot{\mathcal{J}}_{\rm Rey}$ and $ \dot{\mathcal{J}}_{\rm adv} $ by this method is likely sufficient.

\section{Results}
\label{sec:results}

\begin{table*}
\centering
\begin{tabular}{ccccccccc|cc}
\hline
Model&  $\Omega^{-1}$ (year) & $R_{\rm Hill} (R_\odot)$ & $\rho_{0}$ (g cm$^{-3}$) & $T_{0}$ (K) & $H_{\rm gas} (R_\odot)$ & $H (R_\odot)$ & $\Pi_{0}$ & $\Sigma$ (g cm$^{-2}$)  & $\dot{M}_{\rm iso}$ & Measured $\dot{M}$ \\\hline
\texttt{T5e4}&  0.05 &  245.3  & $10^{-10}$ & $5\times 10^4$ & 56.2 & 274.6 & 22.9 & 4805.7 & 0.023 &  0.022
\\\hline
\texttt{T3e4}&  -- & --  & -- & $3\times 10^4$ & 43.5 & 106.1 &4.9 & 1857.0 &  0.038 & 0.014
\\\hline
\texttt{T4e4}&  -- & --  & -- & $4 \times 10^4$ &   50.6 & 180.4 &  11.7 & 3136.5 & 0.031& 0.019
\\\hline
\texttt{T6e4}& -- & --  & -- & $6 \times 10^4$ &   61.5 & 391.9 &  39.5 & 6858.9 &0.016 & 0.021
\\\hline
\texttt{T7e4}& -- & --  & -- & $7 \times 10^4$ &   66.5 & 531.0 &  62.7 & 9292.9 &  0.011 & 0.018
\\\hline
\end{tabular}
\caption{List of model names and input parameters used in our simulations. The stellar and SMBH masses are fixed at $M_\star = 50 M_\odot, M_\bullet = 10^8 M_\odot$ and the distance is $r_\bullet = 0.001$ pc. {The vertical optical depth of the disk is roughly $0.34 \Sigma$ in cgs units for electron scattering dominated opacity. We also show the estimated isotropic accretion rates $\dot{M}_{\rm iso}$ compared to measured accretion rates. } Accretion rate unit is $M_\odot$/yr. }
\label{tab:parameters}
\end{table*}

\begin{figure*}
    \centering
    \includegraphics[width=1.0\textwidth]{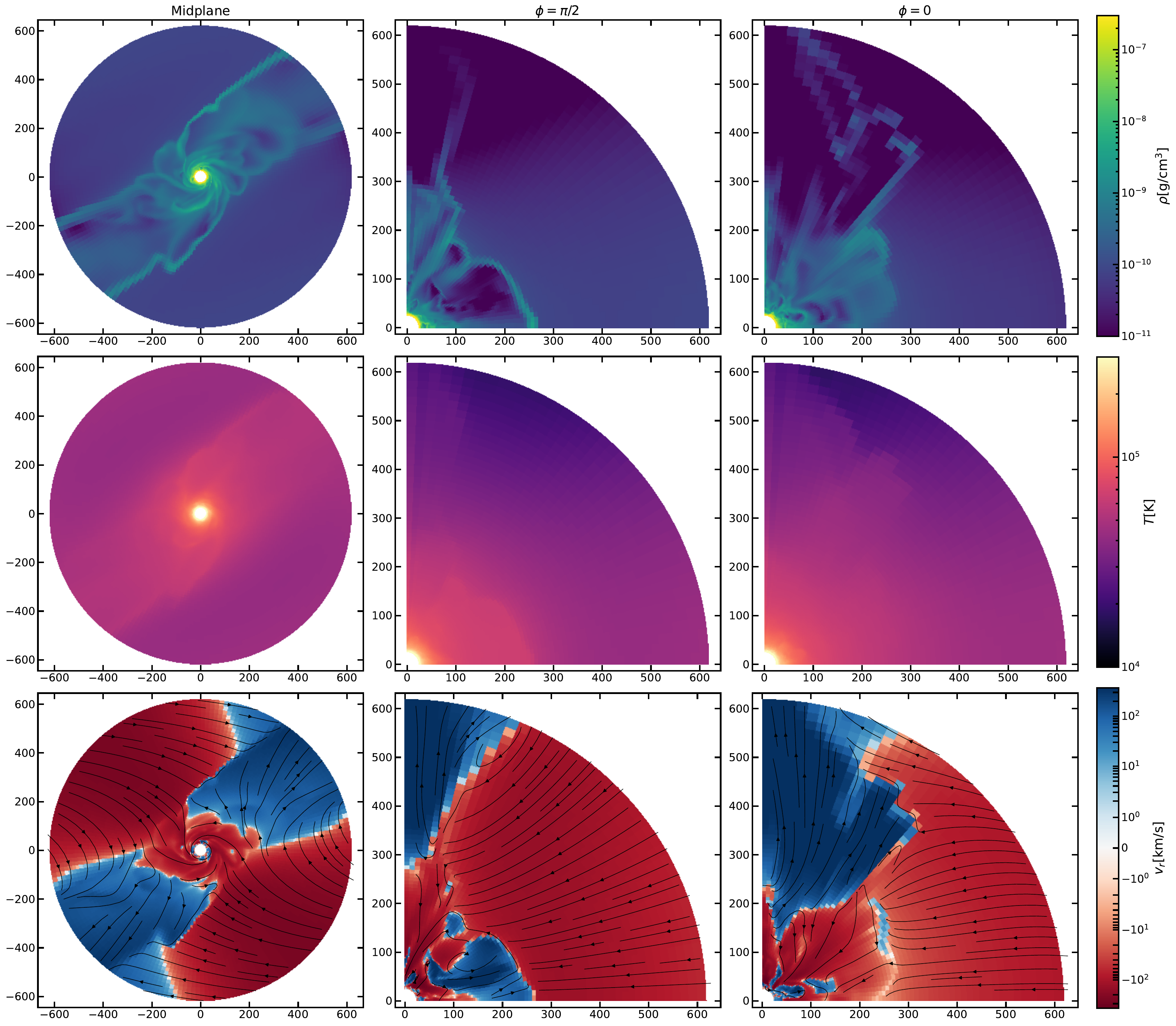}
    \caption{A snapshot of density, temperature and radial velocity distribution for the fiducial run \texttt{T5e4} in quasi-steady state. The projected velocity streamlines are overlaid on the bottom panel.
    The left, 
    middle and right columns correspond to the midplane, 
    $\phi = \pi/2$ (towards host) and $\phi = 0$ 
    (perpendicular to host) vertical distributions. All length units are in $R_\odot$.}
\label{fig:snapshot}
\end{figure*}

\begin{figure*}
    \centering
    \includegraphics[width=1.0\textwidth]{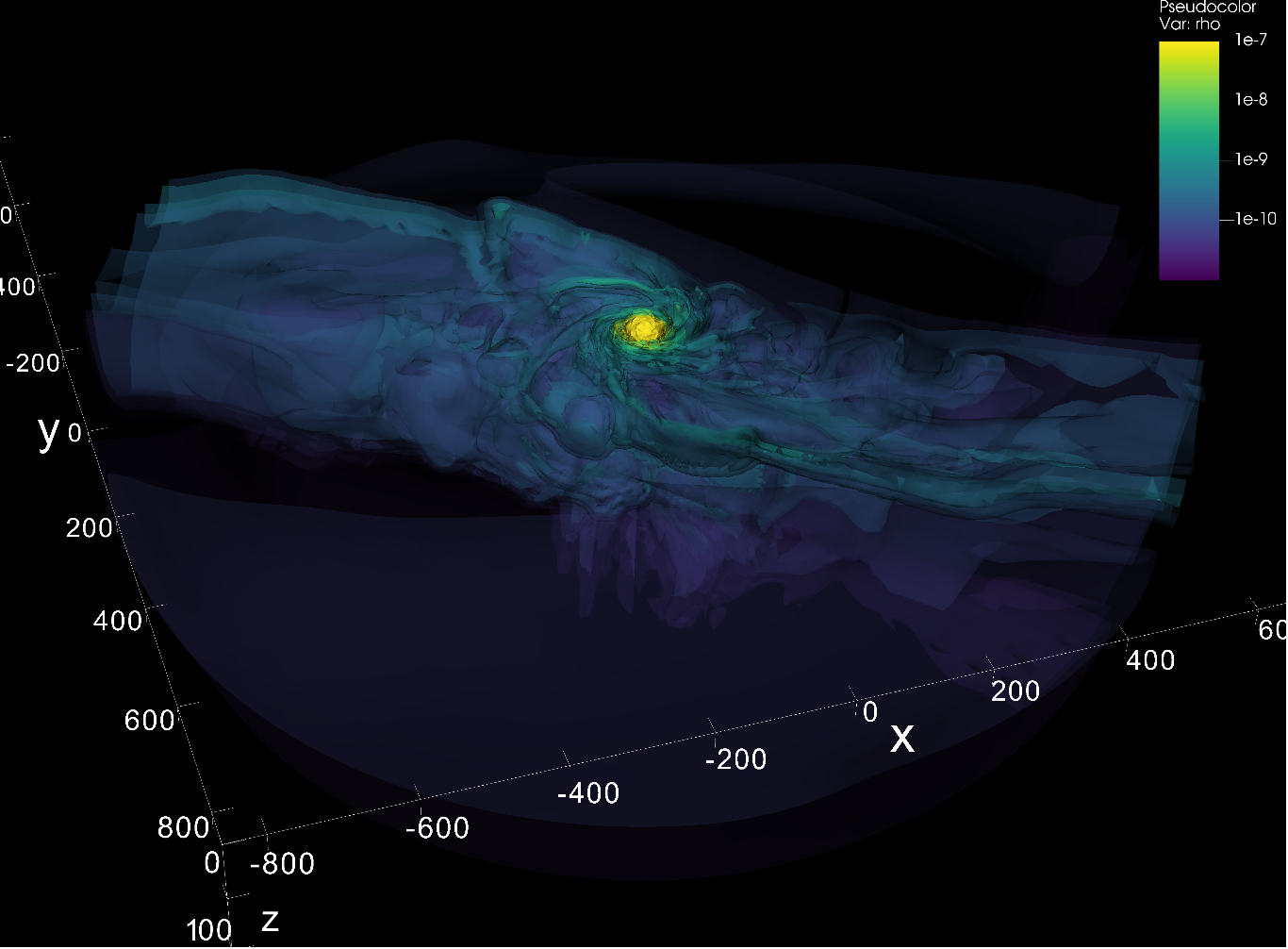}
    \caption{Semi-transparent iso-density contours for the fiducial case 
    showing the 3D structure of high density stellar envelope and density waves. 
    The isodensity contours range from $10^{-11}$ to $10^{-7}$ g/cm$^3$ in physical units. For visualization purpose,  we are viewing the simulation domain from a bottom-up perspective. All length units are in $R_\odot$. Significant density peaks can be seen along the spiral density waves. For a video showing time evolution of iso-density contours towards steady-state see \href{https://yi-xian-chen.github.io/videos/embedded_3Devolution.mp4}{this link}. }
\label{fig:3D_render}
\end{figure*}

\subsection{Fiducial setup}

We first present results from 
our fiducial run with disk midplane temperature $T_0 = 5\times 10^4$ K and density $\rho_0 = 10^{-10}$g/cm$^3$. 
Due to high computational expenses, 
we run the simulation to 1 disk orbital timescale $(2\pi \Omega^{-1})$ 
which takes about 1 million CPU hour, 
when the accretion structure has reached a quasi-steady state. 
\footnote{{although it's far from sufficient to capture long-term global-scale dynamical effects like gap-opening \citep{Lin1986,Li2023} and torque saturation \citep{Masset2001}, 
see \S \ref{sec:long_term}.}}. 
Figure \ref{fig:snapshot} shows typical snapshots of 
$\rho, T, v_r$, 
and also along the 
$\theta = 0$ and $\theta = \pi/2$ vertical cross-sections, 
while Figure \ref{fig:3D_render} shows iso-density contours for this snapshot to illustrate the 3D structure more clearly.


Within the midplane, we observe strong density waves driven by tidal interaction, 
launching off from the Hill radius $\sim 250 R_\odot$. 
While these density waves generally carry materials outwards 
(see dark blue patches in the lower left panel of Figure \ref{fig:snapshot}, 
discontinuous in $\phi$) from the stellar surface faster than the background shear velocity, 
flow in other azimuths are carrying materials towards the star, 
including within the $\phi = 0$ and $\phi = \pi/2$ slices. 
This midplane flow asymmetry is qualititatively similar to that of around a planet embedded within a protoplanetary disk \citep[][see their Figure 3]{Li2023}. 
A significant level of turbulence is present within the density waves, 
consistent with the fact that they form a kind of 
extended convective region with entropy similar to that of the stellar convective zone, 
separated from the background entropy. 


\begin{figure*}
    \centering
    \includegraphics[width=0.45\textwidth]{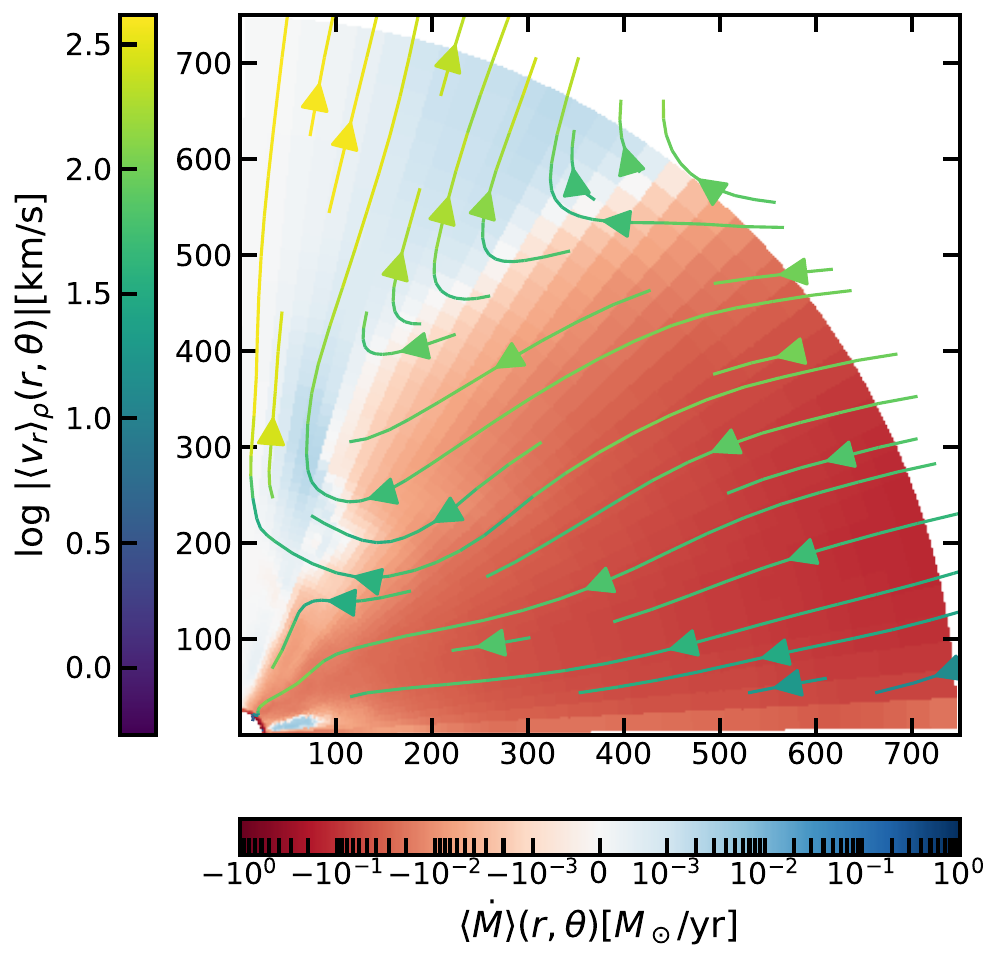}
    \includegraphics[width=0.5\textwidth]{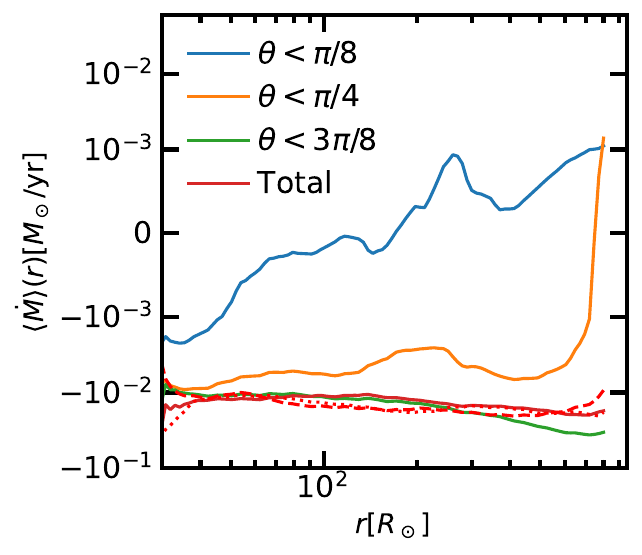}
    \caption{Left panel: 
    time (over 100 snapshots within the final $\Omega^{-1}$) and azimuthally averaged 2D velocity streamlines and accretion rate $\langle\dot{M} \rangle(r, \theta)$ profiles, for fiducial simulation \texttt{T5e4}. Color of streamline indicate the velocity magnitude. Right panel: $\langle\dot{M} \rangle(r, \theta)$ averaged over different polar angle ranges, with red line being the total average. 
    The red dashed and dotted lines indicate averages over the final 50 and 30 snapshots respectively, 
    to show that we measure $\dot{M}$ robustly in a quasi-steady state. }
\label{fig:Mdot_fiducial}
\end{figure*}

For further analysis, 
we take the time-averaged 3D data and create azimuthally-averaged 2D accretion rate and velocity profiles to check their altitudinal dependence. 
By plotting $\langle \dot{M} \rangle(r,\theta)$ distribution and $\langle v_{\rm r} \rangle_\rho (r,\theta)$ streamlines in the left panel of Figure \ref{fig:Mdot_fiducial}, 
we see that on average 
accretion occurs close to the midplane while outflows occur in the polar region. 
The right panel of Figure \ref{fig:Mdot_fiducial} plots the average accretion rate $\langle \dot{M} \rangle(r)$ integrated from from pole to $\theta =\pi/8, \pi/4, 3\pi/8, \pi/2$ respectively, 
which shows that while outflow dominates the region
outside the stellar surface close to the pole (blue line), 
as we include lower altitudes (integrate up to larger $\theta$)
inflow from region close to the midplane eventually dominates and the sum
converges to a quasi-steady net accretion rate of $\dot{M} \sim 2\times 10^{-2} M_\odot$/yr \footnote{henceforth, following the convention of existing literature, we refer to scalar quantity $\dot{M}$ as the radial average of  $ |\langle \dot{M} \rangle(r)|$ despite our definition of mass flux giving $\langle \dot{M} \rangle(r) < 0$}. 
We also plot total $\langle \dot{M} \rangle(r)$ averaged over shorter timescales (the final $0.2\Omega^{-1}$, $0.5\Omega^{-1}$) in dashed and dotted lines to show it converges in a quasi-steady state. 

This vertical anisotropy is closely 
linked to the distribution of radial diffusive radiation flux under the influence of disk geometry. 
In the fully isotropic case, 
a steady-state accretion rate $\dot{M}$ introduces an additional source of luminosity which, 
combined with stellar intrinsic luminosity, 
reduces the gravity of the star isotropically by a factor of $\lambda < \lambda_{\rm crit} = \lambda_{\rm diff}(R_{\rm crit}) < 1$ 
that is self-consistently determined by requiring density and temperature 
at $R_{\rm crit}$
to be similar to boundary values. 
For clarification, we denote the critical radius calculated under the isotropic assumption as $R_{\rm crit, iso}$ 
\citep[][see \S 3.2 therein]{Chen2024} as a function of $\rho_0, T_0$, to distinguish it from the inherently 
anisotropic $R_{\rm crit} (\theta)$ surfaces measured in our simulation and $\dot{M}_{iso}$ as the nominal isotropic accretion rate calculated from $\dot{M}_{\rm iso} = 4\pi \rho_0 R_{\rm crit, iso}^2 $. The explicit expression for $\dot{M}_{\rm iso}$ is given in Equation \ref{eqn:Mdotiso}.

\begin{figure}
    \centering
\includegraphics[width=0.48\textwidth]{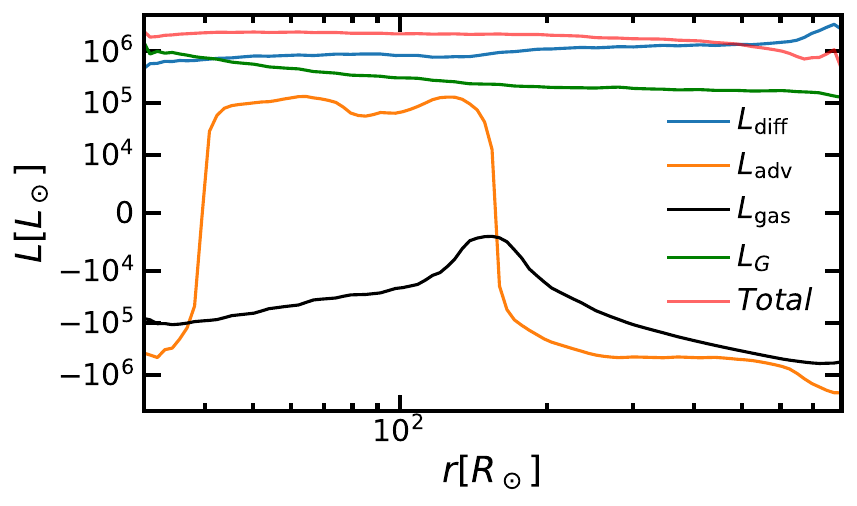}
\caption{average radial profiles for different energy fluxes in the
fiducial run \texttt{T5e4}. Definition of these variables and details of averaging see \S
\ref{sec:diagnostics}.}
\label{fig:energy_fiducial}
\end{figure}

\begin{figure*}
    \centering
    \includegraphics[width=1.0\textwidth]{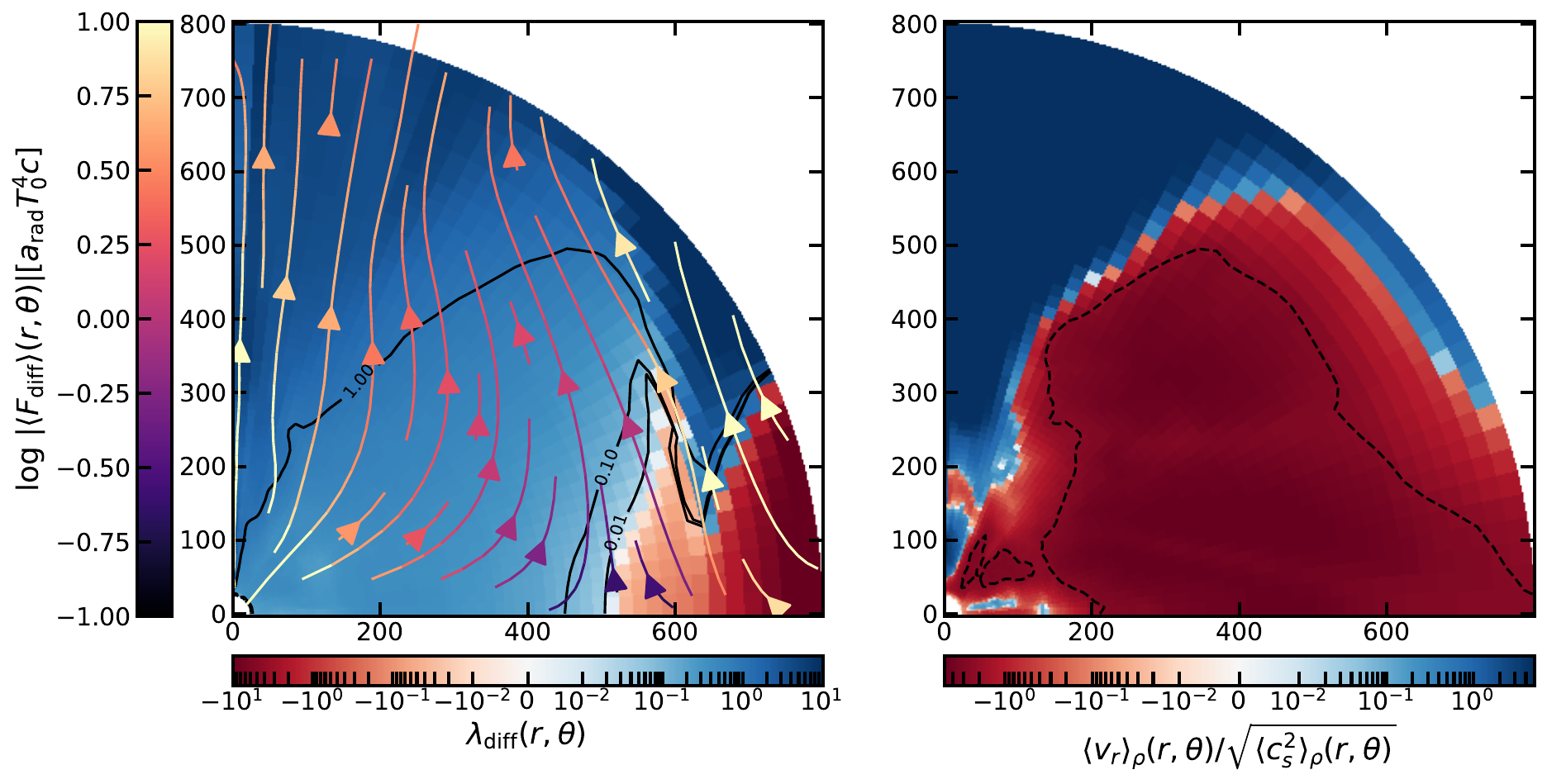}
    \caption{Left panel: 
    time and azimuthally averaged 2D diffusive radiation flux streamlines and $\lambda_{\rm diff}(r, \theta)$ profiles, 
    where $\lambda_{\rm diff}(r, \theta)$ indicates the fraction of radial gravity that radiation can effectively reduce, for fiducial simulation \texttt{T5e4}. Solid black lines are contours of constant $\lambda_{\rm diff}(r, \theta)$. Right panel: 
    ratio of azimuthally averaged radial velocity and root-mean-square gas sound speed. 
    Dashed line shows the $\langle v_r\rangle_{\rho}/\langle c_s^2 \rangle_{\rho}^{1/2}=-1$ contour 
    within which material infall becomes supersonic. }
\label{fig:Rcrit}
\end{figure*}

On an average sense, 
this general conclusion still holds in our anisotropic setup. 
We plot the angle averaged energy flows in Figure \ref{fig:energy_fiducial}, 
which shows clearly that gravitational and advection luminosity are both decreasing inwards, 
and therefore converted to 
released diffusive luminosity that adds on to the stellar luminosity (blue solid line bending up outwards in order to keep total luminosity constant in a quasi-steady state). 
{This is similar to Figure 4 of \citet{Chen2024}, 
only that the ``plateau" of advective luminosity - where $\langle P_{\rm rad} v_r\rangle $ is allowed to be positive despite a negative $\langle  v_r\rangle_{\rho}$ - is more extended. 
This may be due to 
a correlation between radiation pressure and inflow velocity in asymmetric spiral modes absent in isotropic cases, which contributes in addition to pure convective transport.} 
Nevertheless, 
it's the diffusive luminosity that ultimately controls the feedback. 
When we plot azimuthally averaged $\lambda_{\rm diff} (r, \theta)$ and diffusive flux streamlines 
in the left panel of Figure \ref{fig:Rcrit}, 
we can see that at a given radius,  $\lambda_{\rm diff}$ can be relatively small close to the midplane, 
but then gradually increases towards the pole and eventually 
exceeds unity at the pole. 
This suggests that much of the accretion luminosity is released near the pole, 
where it drives super-Eddington outflows. 
In contrast, 
in the relatively optically thick midplane, 
the radiation field is primarily determined by vertical components from the temperature gradient of the background disk, 
and the radial feedback from stellar luminosity is effectively suppressed 
(or ``deflected" into higher altitudes, 
as seen from the flux streamlines in the left panel of Figure \ref{fig:Rcrit}). 

The outcome of this asymmetry is the establishment of an anisotropic critical surface/accretion cross section $R_{\rm crit}(\theta)$, 
plotted as the $\langle v_r\rangle_{\rho}/\langle c_s^2 \rangle_{\rho}^{1/2}=-1$ contour in the right panel of Figure \ref{fig:Rcrit}. 
This boundary reaches the outer edge of simulation domain at angles close to midplane (minimal feedback), 
but gradually circles around towards $\sim 4-600 R_\odot$ at higher altitudes $(\theta < 60^\circ)$, 
until at $\theta < 30^\circ$ where $\lambda_{\rm diff}$ reaches $\gtrsim$1, 
the inflow transitions to outflow. 
In summary, 
compared to the isotropic case, 
radiative feedback is enhanced at the pole and reduced in the midplane. 
These competing effects collectively 
yield the remarkable outcome that the measurement of accretion rate $\dot{M} \sim 2\times 10^{-2} M_\odot$/yr 
is quite consistent with what we would predict in the isotropic framework, 
using the estimate of \citet{Chen2024}. 

\subsection{Varying the Disk Scale Height}

Relative to the fiducial parameters, 
we perform a series of simulations with different $T_0$ (see Table \ref{tab:parameters}) 
and disk scale heights, 
and record all their time-average profiles as well as accretion rates. 
This allows us to systematically assess how disk geometry, from thick to thin, 
influences deviations from isotropic stellar accretion.

To compare different geometries, we adopt the terms “subthermal” and “superthermal” from planetary accretion contexts but generalize their meanings \citep{Li2021, Li2023, Choksi2023}. 
If there is
(i) minimal radiative feedback and
(ii) negligible radiation pressure,
the relationship among the embedded stellar object's Bondi radius $R_B = GM_{\star}/c_{\rm s, gas}^2$, 
Hill radius $R_{\rm Hill} = (q/3)^{1/3} r_\bullet$, 
and scale height $H = H_{\rm gas}$ is solely determined by the mass ratio $q = M_\star/M_{\bullet}$:

\begin{itemize}
\item When $q \lesssim (H_{\rm gas}/r_\bullet)^3$, 
such that $R_B \lesssim R_{\rm Hill} \lesssim H_{\rm gas}$, 
accretion occurs in a deeply embedded, quasi-3D regime with an accretion cross-section set by $\sigma \sim R_B^2$. 
This scenario is referred to as subthermal accretion and accretion rate is set by $\dot{M}_{B}:= 4\pi \rho c_{\rm s, gas} R_B^2$.  
\item when $q \gtrsim (H_{\rm gas}/r_\bullet)^3$, 
such that $R_B \gtrsim R_{\rm Hill} \gtrsim H_{\rm gas}$, 
the companion's sphere of influence extends above the disk, and accretion becomes more 2D, 
with a cross-section set by $\sigma \sim R_{\rm Hill} H_{\rm gas}$. This scenario is referred to as superthermal accretion and the Hill accretion rate is set by $\dot{M}_{\rm Hill} =  4\pi \rho H_{\rm gas} \Omega R_{\rm Hill}^2 = 4\pi \rho c_{\rm s, gas} R_{\rm Hill}^2$ since the impact velocity is $\sim \Omega R_{\rm Hill} > c_{\rm s, gas}$.
\end{itemize}

In a high-temperature environment where radiation effects are non-negligible, 
the disk scale height is enhanced to $H = H_{\rm gas}\sqrt{1 + \Pi_0}$ which could be much larger than $H_{\rm gas}$. 
Also, 
the nominal 3D accretion radius $R_{\rm crit, iso}$ is modified from $R_B$ by an effective gravity factor which also depends sensitively on disk and stellar parameters. 
To be specific, following \citet{Chen2024}, the nominal accretion rate, gravity reduction and critical radius ($\dot{M}_{\rm iso}$, $\lambda_{\rm diff, iso}$, $R_{\rm crit, iso}$) are connected by 

\begin{equation}
\begin{aligned}
\lambda_{\rm crit, iso}&=\lambda_{\star}+\frac{\dot{M}_{\rm iso}}{L_{\mathrm{Edd}}}\left( \frac{G M_{\star}}{R_{\star}}+ \frac{a T_{\mathrm{out}}^4}{\rho_{\mathrm{out}}}\right)
\\
R_{\rm {crit,iso}}&=\left(1-\lambda_{\rm {crit,iso}}\right) R_{\rm B}    
\\
\dot{M}_{\rm {iso}} &= \left(1-\lambda_{\rm {crit,iso}}\right)^2 \dot{M}_{\rm B}
        \end{aligned}
        \label{eqn:Mdotiso}
\end{equation}

As a result, 
the hierarchy among these characteristic radii becomes a non-linear function of relevant parameters.
More importantly, 
even after accounting for isotropic feedback effects, 
we have shown that $R_{\rm crit}(\theta), \lambda_{\rm diff}(r, \theta)$ can both be quite 
anisotropic across polar angles, 
making a straightforward definition of effective accretion radius impractical. 
E.g. in the fiducial case, 
$R_{\rm Hill}$, $H_{\rm gas}$, as well as $R_{\rm crit, iso}\sim 300 R_\odot$ (pre-calculated from assuming isotropic accretion as in \citet{Chen2024}) are comparable, 
placing the system in a marginally superthermal regime. 
However, 
Figure \ref{fig:Rcrit} shows that the anisotropic $R_{\rm crit}(\theta)$ surface
can reach out to $500 R_{\odot}$ near the midplane, 
while in outflowing polar regions, 
it does not exist at all.

\begin{figure*}
    \centering
    \includegraphics[width=0.48\textwidth]{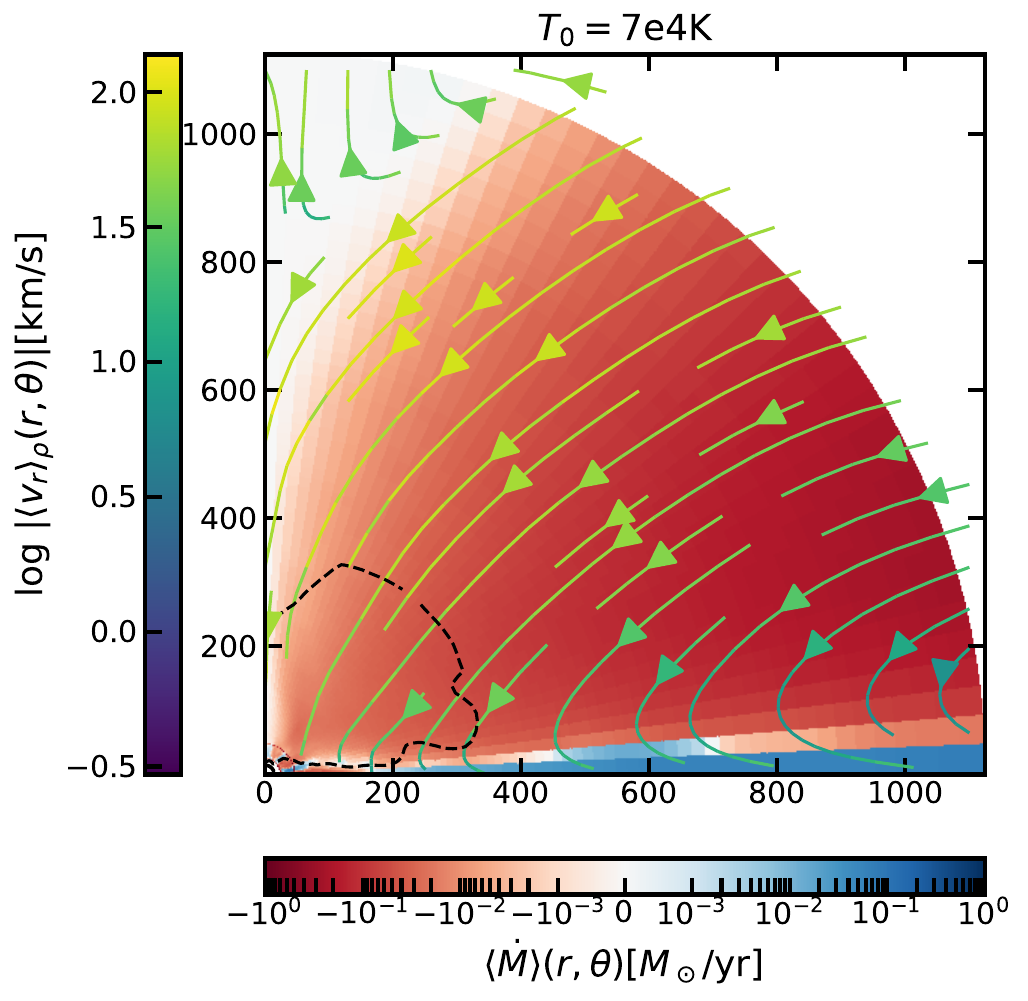}
\includegraphics[width=0.495\textwidth]{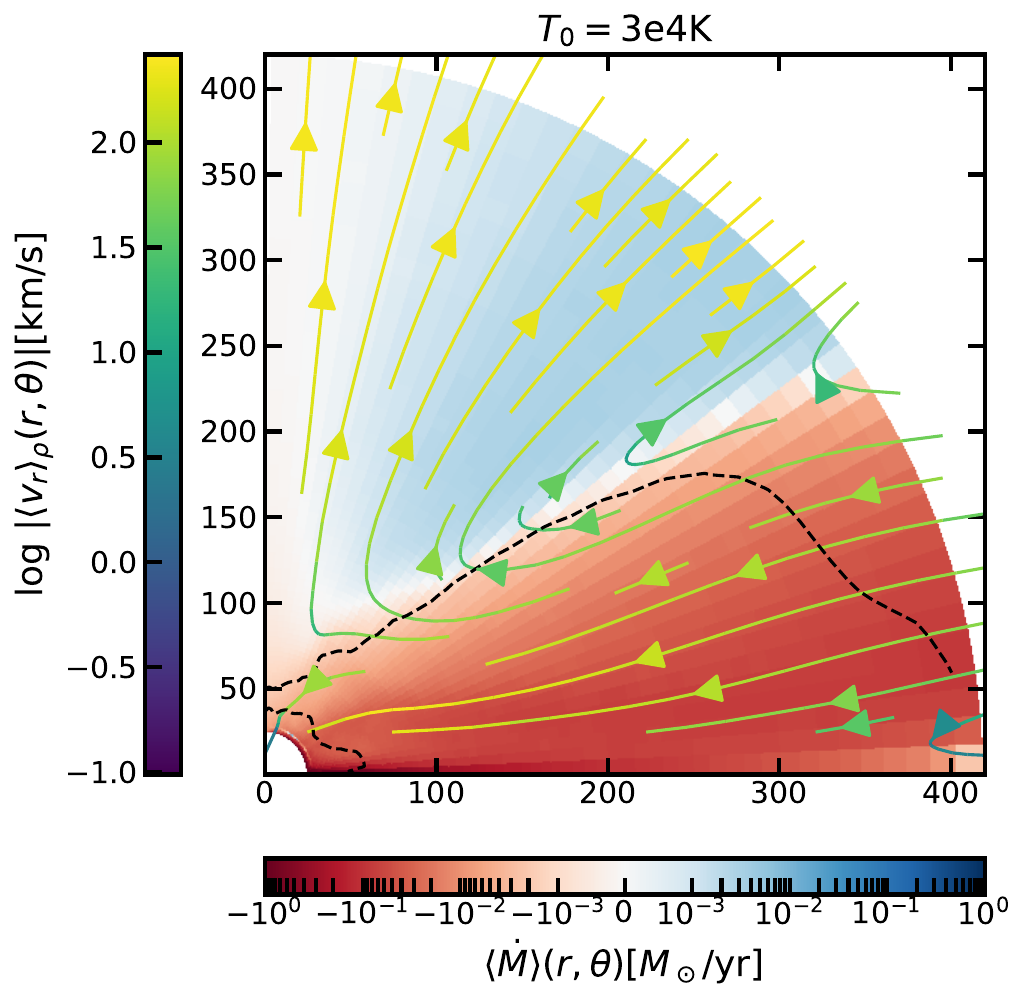}
    \caption{Similar to Figure \ref{fig:Mdot_fiducial}, but for the representative superthermal (left panel) case \texttt{T3e4} and subthermal (right panel) case \texttt{T7e4}. The $\langle v_r\rangle_{\rho}/\langle c_s^2 \rangle_{\rho}^{1/2}=-1$ contour is indicated as black dashed lines, as in Figure \ref{fig:Rcrit}.}
\label{fig:varying_scaleheight}
\end{figure*}

Nevertheless, 
since $R_{\rm crit, iso}$, 
the only quantity that we can pre-calculate from Equation \ref{eqn:Mdotiso}, 
remains a steeply decreasing function of temperature while $H$ sharply increases with $T_0$ 
(especially when radiation pressure dominates), 
the condition $R_{\rm crit, iso} < R_{\rm Hill} < H$ easily holds at high $T_0$ and we can still generally identify this as the ``hot" or subthermal regime. 
The self-consistency of this regime lies in the observation that as accretion becomes more isotropic, 
$R_{\rm crit}(\theta)$ is more well-defined and converges towards $R_{\rm crit, iso}$. 
The left panel of Figure \ref{fig:varying_scaleheight} shows azimuthally averaged accretion rate profiles for
a typical subthermal case \texttt{T7e4}  ($T_0 = 7\times 10^4$K) with a much larger scale height compared to the fiducial case $H = 531 R_\odot$. 
As expected, 
accretion is more isotropic 
with most solid angles permitting inflow, 
and $R_{\rm crit}(\theta)$ appears more spherical and consistent 
with the isotropic estimate of $R_{\rm crit, iso} \sim 200 R_\odot$ 
along most solid angles. 


Going down to low enough temperature, the condition $R_{\rm crit, iso} > R_{\rm Hill} > H$ can generally be satisfied with other hierarchies occupying only a narrow temperature range. 
Although $R_{\rm crit}(\theta)$ is expected to deviate strongly from $R_{\rm crit, iso}$ within this relevant parameter space,  
the boundary of this regime can still be broadly interpreted as the transition to the superthermal or ``cold" accretion scenario.
For example, 
in the typical superthermal case \texttt{T3e4} ($T_0 = 3\times 10^4$ K), shown in
the right panel of Figure \ref{fig:varying_scaleheight}, we see that most accretion is confined to the solid angle occupied by the thin disk, 
while outflows — carrying minimal mass flux due to the low-density gas — occur above the disk surface. 
This geometric stratification restricts the effective solid angle for accretion, 
making the calculation of $R_{\rm crit, iso}$ less physically meaningful. 
However, this distinction becomes irrelevant, 
as the accretion cross-section is now primarily determined by $R_{\rm Hill}$ and $ H$ rather than $R_{\rm crit, iso}$. 


\begin{figure}
    \centering
    \includegraphics[width=0.48\textwidth]{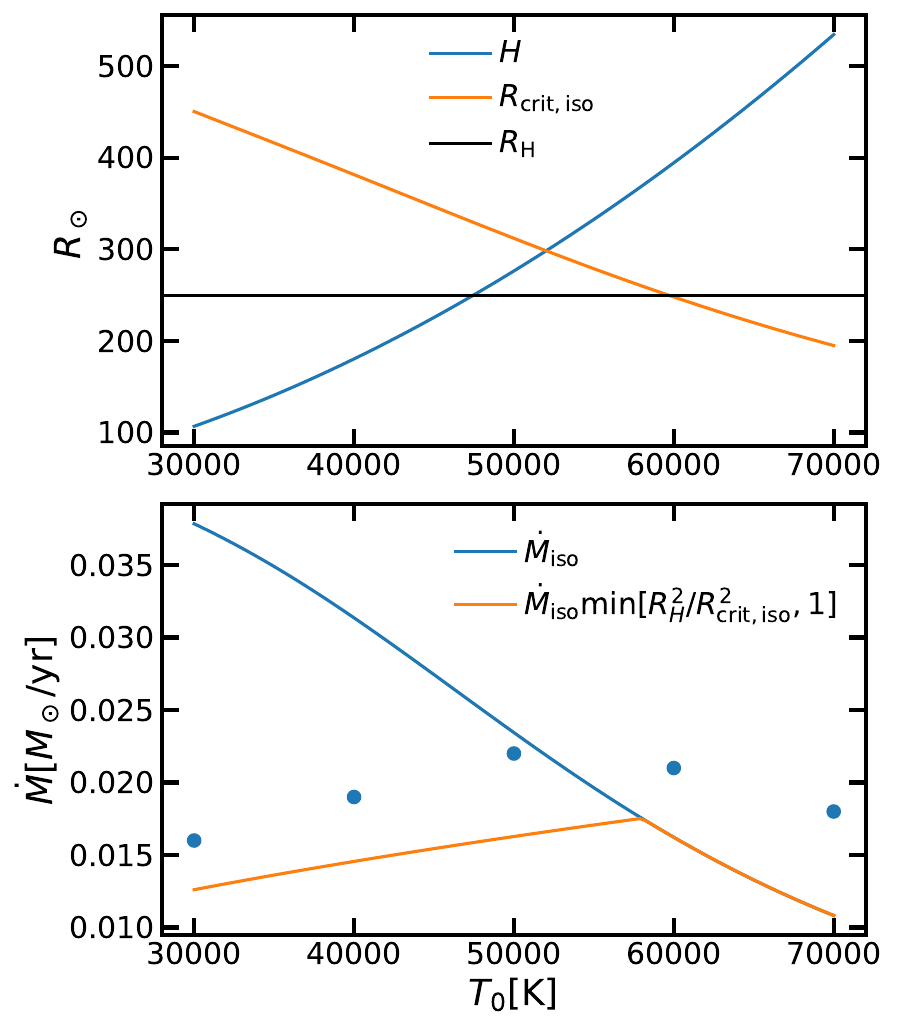}
    \caption{Top panel: hierarchy between $H, R_{\rm crit, iso}$ and $R_{\rm Hill}$ for our simulation $\rho_0, r_\bullet, \Omega$ parameters across a range of $T_0$. Since $H, R_{\rm crit, iso}$ are steep functions of temperature, the superthermal regime dominates at low $T_0$ and subthermal regime dominates at high $T_0$. Lower panel: accretion rates measured from our simulations (blue dots), plotted against estimates from assuming isotropic accretion (blue line) and taken into account possible reduction of effective accretion cross section (orange line).}
\label{fig:mdot_summary}
\end{figure}

The upper panel of Figure \ref{fig:mdot_summary} shows $H$, $R_{\rm crit, iso}$, and $R_H$ for our simulation parameters $\rho_0$, 
$r_\bullet$, and $\Omega$ across a range of $T_0$. Given the steep temperature dependence of $H$ and $R_{\rm crit, iso}$, 
we confirm that $H\lesssim R_{\rm Hill} \lesssim R_{\rm crit, iso}$ for $T_0 \lesssim 5 \times 10^{4}$ K, $H \gtrsim  R_{\rm Hill}  \gtrsim  R_{\rm crit, iso}$ for $T_0 \gtrsim 5 \times 10^{4}$ K. 
In the lower panel, 
we plot the final mass accretion rates from our simulations as a function of $T_0$ as well as two different estimates. 
In the subthermal regime, 
the measured accretion rates converge smoothly toward the isotropic estimate $\dot{M}_{\rm iso}$. 
In contrast, 
the superthermal regime exhibits a marked transition, with accretion rates significantly lower than $\dot{M}_{\rm iso}$, 
settling into a new scaling that's consistent with a $\propto R_{\rm Hill}^2$ trend. In summary, a useful prescription that captures the high $T_0$ and low $T_0$ limits would be

\begin{equation}
    \dot{M} \sim 4\pi \rho_0 c_{\rm s, gas, 0} \min[R_{\rm crit, iso}^2, R_{\rm Hill}^2] 
    \label{eqn:ratescaling}
\end{equation}

without feedback, 
this marks a transition from $\dot{M} \propto M_\star^2$ towards $\dot{M} \propto M_\star^{2/3}$ 
as 
stellar mass grows sufficiently large, 
similar to the planet context \citep{Choksi2023,Li2023}. 
Accounting for feedback, it be calculated from Equation \ref{eqn:Mdotiso} 
that when $M_\star$ is sufficiently large and gravity feedback term dominates, 
the isotropic accretion rate converges to the stellar Eddington rate

\begin{equation}
    \dot{M}_{\rm iso} \rightarrow \dot{M}_{\rm Edd} = \dfrac{4\pi (1-\lambda_\star)R_\star c}{\kappa},
\end{equation}

In this limit, for $1-\lambda_\star \sim \mathcal{O}(1)$, 
$\dot{M}_{\rm iso}$ may still follow a relatively steep power dependence on $M_\star$ if $R_\star \propto M_\star^{0.8}$ at first, 
before a mild flattening towards $\dot{M} \propto M_\star^{2/3}$ for even larger $M_\star$ in the superthermal regime. In addition, gap-opening effect may further reduce the accretion rate on a longer timescale \citep{Lin1986}.
However, 
if $\lambda_\star$ rapidly approaches unity for $M_\star \gtrsim 100M_\star$, 
the dependence of $\dot{M}$ on stellar mass can become highly non-linear. 
If $R_{\rm crit, iso}$ 
becomes as small as the stellar radius due to significantly reduced gravity, 
transition to the superthermal regime may be impeded, although the occurence of adiabatic accretion, 
when the background density is high enough, 
could still result in runaway accretion 
under effect of the stellar envelope's self-gravity. We offer a more detailed discussion on effects of gap-opening and/or adiabatic accretion in \S \ref{sec:discussions}.

\subsection{Angular Momentum Budget}
\label{sec:AMtransport}

One key difference between our circumstellar accretion flow 
and simulation results of circumplanetary disks (CPDs) is the vertical velocity field.
For planets embedded in a protoplanetary disk, 
inviscid or low-viscosity simulations show that a rotational CPD typically forms within $\sim 0.5 R_{\rm Hill}$ 
\citep{Choksi2023,Li2023}, 
although the degree of rotation support 
(A.K.A proximity to Keplerian) depends on the mass ratio. 
Notably, 
in their simulations, 
material typically spiral \textit{outwards} at low altitudes, 
feeding a polar accretion inflow. 
The ``angular momentum barrier" 
argument for this meridional flow to establish for CPDs is that, 
in the absence of convection or active turbulence, 
if the Reynolds stress is small due to 
low effective viscosity, 
accretion will have to proceed from the pole 
to keep $\dot{\mathcal{J}}_{\rm adv}$ minimal 
(such that accreted materials have minimum $v_\phi$) 
for $\dot{\mathcal{J}}_{\rm Rey} + \dot{\mathcal{J}}_{\rm adv} \sim {\rm Const}$ to be satisfied. 
In such framework, 
material at high altitudes is preferentially accreted because it carries less angular momentum 
than midplane material.

However, 
what we observe from our azimuthally averaged flow patterns is mostly the opposite, 
highlighting the importance of radiative feedback. 
The anisotropic reduction of effective gravity significantly alters the accretion structure, 
overriding the classic angular momentum barrier effect. 
This is more clearly shown in our fiducial and superthermal case \texttt{T3e4}. 
Case \texttt{T7e4} allows for a small solid angle of outflow near the midplane, 
possibly 
indicating convergence toward CPD results 
as anisotropy diminishes.

\begin{figure}
    \centering
    \includegraphics[width=0.45\textwidth]{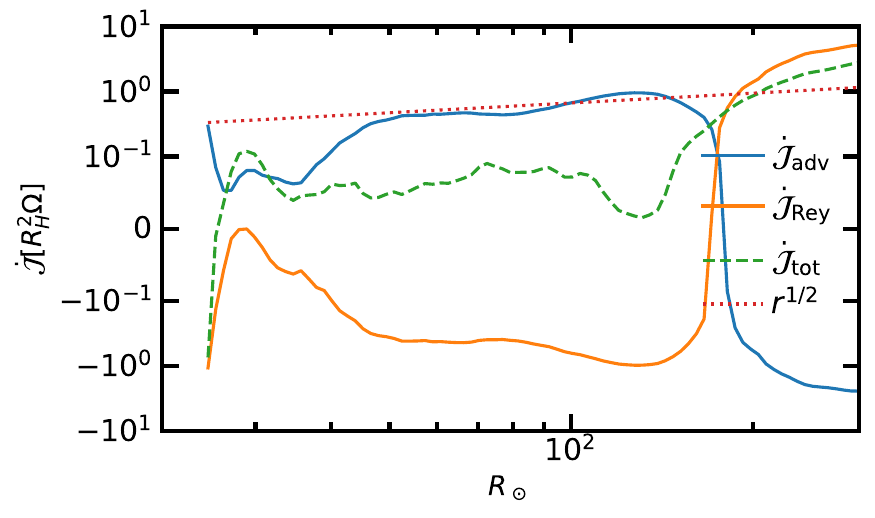}
    \includegraphics[width=0.45\textwidth]{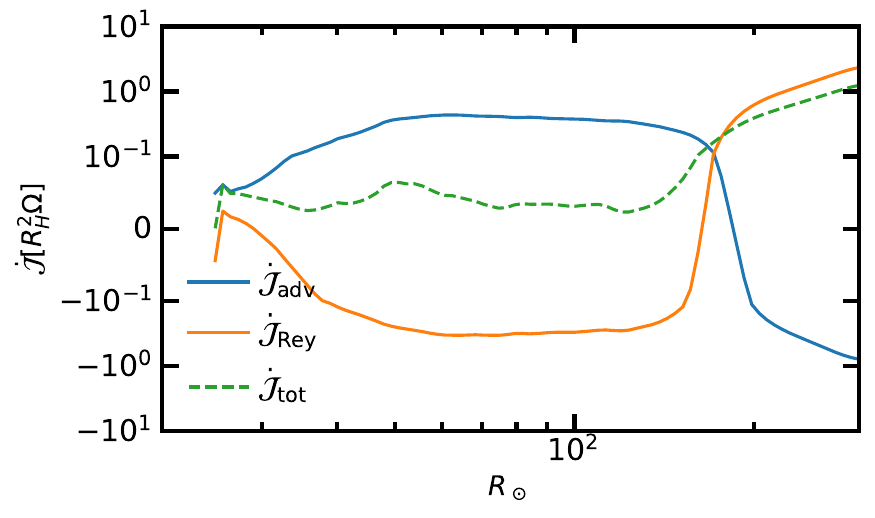}
    \caption{Radial angular momentum fluxes defined in \S \ref{sec:diagnostics} 
    for cases \texttt{T3e4} and \texttt{T7e4}. 
     Green dashed lines show the sum $\dot{\mathcal{J}}_{\rm tot} = \dot{\mathcal{J}}_{\rm Rey} + \dot{\mathcal{J}}_{\rm adv}$, which is expected to be a small constant along the accretion flow, 
     although averages in the inner stellar envelope within $\sim 40 R_\odot$ 
     may be subject to uncertainties introduced by stellar convection. }
\label{fig:varying_scaleheight_Jdot}
\end{figure}

For an accretion profile with non-negligible $\dot{\mathcal{J}}_{\rm adv}$ to be established, 
Reynolds stress should be responsible for transporting angular momentum \textit{outwards} near the midplane. 
In Figure \ref{fig:varying_scaleheight_Jdot}, 
we plot $\dot{\mathcal{J}}_{\rm adv}$ 
and $\dot{\mathcal{J}}_{\rm Rey}$ profiles (defined in \S \ref{sec:diagnostics}) 
for two cases representing the superthermal and subthermal regimes, normalized by $R_{\rm Hill}^2 \Omega$. 
Within $R_{\rm Hill}$, 
the Reynolds stress from turbulence effectively transports angular momentum outward in both regimes, 
counteracting the inward transport by the positive advection term. 
This balance ensures that $\dot{\mathcal{J}}_{\rm Rey}$ and $ \dot{\mathcal{J}}_{\rm adv}$ nearly cancels out within the radial range 
from the Hill radius $\sim 200 R_\star$ down to deep within the stellar envelope, yielding a $\dot{\mathcal{J}}_{\rm tot}$ much lower in magnitude. 

One can also clearly see the quantitative difference between the rotational aspect of accretion flow in \texttt{T3e4} and \texttt{T7e4} from the different trends in $\dot{\mathcal{J}}_{\rm Rey},  \dot{\mathcal{J}}_{\rm adv}$. 
For \texttt{T3e4}, 
the circum-stellar disk around the superthermal companion has a nearly Keplerian azimuthally-averaged $v_\phi$ profile, 
therefore
the negative value 
$\dot{\mathcal{J}}_{\rm adv} \sim  v_\phi  r \sim \sqrt{GMr}$ 
decreases towards the center $\propto r^{1/2}$
(the magnitude of $\dot{\mathcal{J}}_{\rm Rey}$ also changes correspondingly). 
Effectively, 
this implies the circularization or truncation radius of circumstellar flow to be constrained 
by $\sim R_{\rm Hill} < R_{\rm crit, iso}$ when the Hill radius constrains the horizontal effective accretion cross section, 
consistent with $\dot{\mathcal{J}}_{\rm adv} \sim R_{\rm Hill}^2 \Omega$ at $\sim R_H$.

Similarly to CPD studies 
\citep[e.g.][]{Sagynbayeva2024}, the circumstellar disk in the sub-thermal case of \texttt{T7e4} 
is less rotationally supported than the super-thermal case, 
resulting in a more isotropic flow. 
Although there is some midplane outflow, $\dot{J}_{\rm adv}$ is still generally positive due to inflow from regions near the midplane, although not as strong as case \texttt{T3e4}. 
Midplane rotation becomes Keplerian (or in other words, $\dot{\mathcal{J}}_{\rm adv}$ steepens to nearly $\propto r^{1/2}$) only within 
a circularization radius constrained by some circularization radius $R_{\rm circ} <  R_{\rm Hill}$. 
Consequently, 
the $\dot{\mathcal{J}}_{\rm adv}$ and $\dot{\mathcal{J}}_{\rm Rey}$ profiles remain flat in the ``ballistic'' region of the flow beyond the critical radius. 
Note that the convergence to Keplerian rotation at $R_{\rm circ}$ 
aligns with a flatter $\dot{\mathcal{J}}_{\rm adv} \sim \sqrt{GM_\star R_{\rm circ}}$ 
extending out to $R_{\rm Hill}$. 
{Note that in the much more optically thick regime, 
a pressure-supported envelope may form, similar to what is seen in adiabatic simulations of circumplanetary flows \citep{Fung2019}, 
where rotation remains significantly sub-Keplerian down to the companion core, 
in contrast to the convergence to Keplerian seen in Figure \ref{fig:varying_scaleheight_Jdot}.
We will return to this point in \S \ref{sec:slodiff}. }

\begin{figure}
    \centering
    \includegraphics[width=0.45\textwidth]{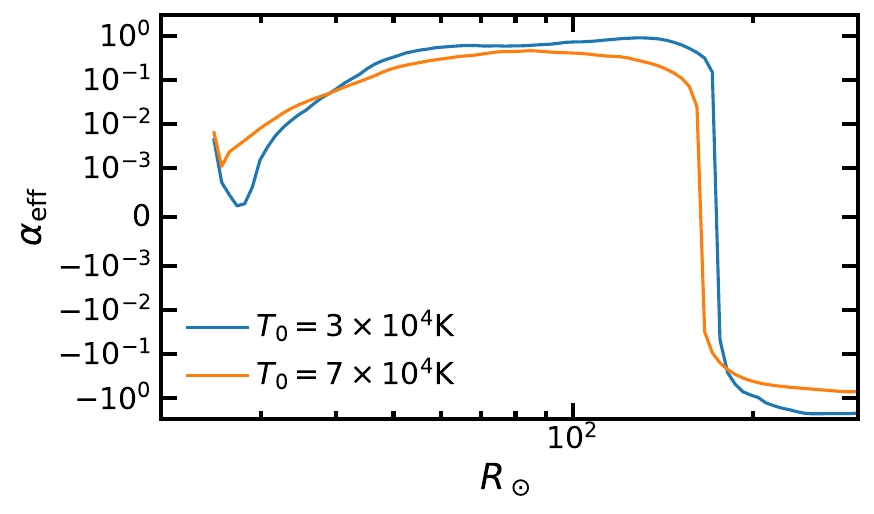}
    \caption{Effective accretion parameter $\alpha_{\rm eff}$ 
    for cases \texttt{T3e4} and \texttt{T7e4}. Both cases yield $\alpha_{\rm eff} \sim 0.1-1$ within $50-200 R_\odot$.}
\label{fig:alpha_comp}
\end{figure}

If we associate Reynolds stress 
in this circum-stellar disk (CSD) with the viscosity parameter $\alpha_{\rm eff}$ as defined in 
\S \ref{sec:diagnostics}, 
we measure $\alpha_{\rm eff}$ 
to be generally within a range of $0.1-1$ 
in the region between $R_H$ and the stellar convective zone $50-200 R_\odot$ for both subthermal and superthermal simulations, 
as shown in Figure \ref{fig:alpha_comp}. 
{Although subject to many uncertainties, this value is also 
broadly consistent with the \textit{requirement} of sustaining the observed accretion rate when modeling the CSD as a nearly-isothermal, 
standard thin disk \citep{SS1973}:}

\begin{equation}
    \dot{M} \approx 3\pi \nu \Sigma_{\rm CSD} = 6\pi \alpha_{ \rm eff} c_s ^3 \Omega_{\rm CSD}^{-2} \rho,
\end{equation}

Where we adopted $\nu  = \alpha_{ \rm eff} c_s^2/\Omega_{\rm CSD}$ and $\Sigma_{\rm CSD} = 2 \rho H_{\rm CSD} =2 \rho c_s/\Omega_{\rm CSD}$. 
Assuming the midplane density and sound speed are comparable to those of the background disk and adopting characteristic values for other parameters, 
we get

\begin{equation}
\begin{aligned}
    \alpha_{ \rm eff} &  \approx 1 \left(\dfrac{\dot{M}}{0.01 M_\odot /{\rm year
    }}\right) \left(\dfrac{\rho_0}{10^{-10}}\right)^{-1}  \left(\dfrac{c_s}{10^7{\rm cm/s}}\right)^{3}\\ & \times \left(\dfrac{\Omega_{\rm CSD}}{50/{\rm year}}\right)^{-2} 
    \end{aligned}
\end{equation}

The characteristic value 
of $50/{\rm year}$ for $\Omega_{\rm CSD}$ 
is estimated from the Keplerian frequency at 200$R_\odot$ from a 50$M_\odot$ star. 

Regarding the angular momentum budget, 
so far we have confirmed that {Reynolds} stress can indeed account for highly efficient angular momentum transport associated with our accretion rates. 
{This alone does not reveal what sustains the $r\phi$ component of the Reynolds stress in the absence of magnetic fields.
One might ask if whether vertical convection or related hydrodynamic instabilities might play a role \citep{Lin+Papaloizou1980, LesurOgilvie2010, Nelson+2013, Lyra+2014, Klahr+Hubbard2014}, 
but none of these mechanisms appears capable of producing the 
transport rates we observe ($\alpha_\mathrm{eff} \gtrsim 0.1$).}

{We therefore consider spiral shocks to be more likely responsible.
It has long been established that spiral shocks excited by the tidal field of a binary companion can drive accretion, even in otherwise laminar disk flows \citep[and references therein]{Sawada+1986,Spruit1987,JuStone2016}.
The fine details of 
how the companion excites the shocks remain obscure \citep{Xu+Goodman2018}, 
but it is agreed that $\alpha_\mathrm{eff}$ is a steep function of the aspect ratio $H/r$, or equivalently, 
the reciprocal of the orbital Mach number $\mathcal{M}=c_s/\bar v_\phi$. 
\cite{Spruit1987} found $\alpha_\mathrm{eff}\propto\mathcal{M}^{-3}$ for his self-similar solutions.
The importance of such shocks for the disks of cataclysmic variables (CVs), where $\mathcal{M}\gtrsim30$ (higher than in most numerical simulations) 
is therefore unclear.
The disks studied here
are much thicker than those of CVs, with $\mathcal{M}\lesssim5$, 
because the Hill radius of the companion is comparable to the thickness of the disk, 
so one may expect spiral shocks to be relatively efficient 
at transporting angular momentum.
While we do not provide a systematic study of this mechanism here, we have made an approximate analysis that suggest $\alpha_\mathrm{eff}\gtrsim0.1 - 1$ is plausible from spiral arms, as detailed in Appendix \ref{sec:appA}.
}

\section{Discussions}
\label{sec:discussions}

\subsection{Long term and global effects}
\label{sec:long_term}

{The runtime of our simulations $(2\pi \Omega^{-1})$ is sufficient to establish a dynamical steady state where the accretion rate profile converges, 
this is because both sound crossing and free-fall timescales within the Hill and the Bondi radius are at most comparable to $\Omega^{-1}$, 
while the radiative diffusion timescale is even shorter (See \S \ref{sec:slodiff} for a discussion of cases where the last condition does not hold). }
However, 
additional long-term effects would become significant if the simulation domain were larger and the integration time extended.

First, our setup does not include the heating source of the background AGN disk. 
In reality, this heating should come from the disk's intrinsic turbulence, driven by gravitational instability (GI) or magneto-rotational instability (MRI), which we do not explicitly model. 
Instead, we impose the background disk as a boundary condition, continuously feeding it into the simulation domain. 
Since gas injected from the outer boundary takes only $\sim \Omega^{-1}$ to reach the star through shear velocity, there is insufficient time for vertical radiation flux to cool it significantly, hence no need for an additional heating to be present.
As a result, 
the disk stream maintains its approximate vertical profile (Equation \ref{eqn:verticalprofile}) until it is accreted onto the star/circumstellar flow. For global simulations with self-consistent boundary conditions, Magnetic fields and/or self-gravity are needed to provide the necessary heating for proper energy balance.
Notably, with star(s) actively accreting and irradiating in the disk midplane, 
stellar luminosity could be an \textit{extra} heating source that contribute to the overall disk emission and extend the effective disk radius \citep{Sirko2003, Thompson2005, ChenLin2024} — either through radiation reprocessing (if buried within the disk, as in subthermal cases) or as direct emission (if the polar regions are sufficiently cleared out, as in superthermal cases). 
Turbulence from the circum-stellar flow can be another thermal heat source.
Multi-wavelength radiative transfer analysis is needed to assess these sources' impact on the disk's total emission/spectral energy distribution, energy budget, 
and the resulting vertical/radial equilibrium temperature profile. 

Second, we have not accounted for the gap-opening effect \citep{Papaloizou1984,Lin1986,Kanagawa2015, Chen2020}, 
which could deplete gas density in the companion’s vicinity due to the stellar companion's Lindblad torque acting on the surrounding disk. 
This process occurs on a much longer viscous timescale and is inherently global, 
requiring more self-consistent boundary conditions to capture accurately \citep{Li2023}. 
Moreover, 
since turbulent viscosity from GI and/or MRI 
— the key factor regulating gap formation — is not modeled in our simulations as we mentioned, 
simply extending the runtime to assess the depletion factor 
would also likely be inconclusive. 
Global disk simulation centered on the SMBH, 
with sink-cell prescriptions for modeling the embedded star, 
is probably needed to study this phenomenon in detail. 
{Moreover, global torques acting on the embedded star may drive its orbital evolution, 
although torque convergence requires modeling of gap opening on the viscous timescale \citep{Lin1986} 
as well as the saturation of corotation torques on the horseshoe synodic timescale \citep{Masset2001}, 
which is beyond the scope of this paper. }

\subsection{Slow Diffusion Regime}
\label{sec:slodiff}

In \citet{Chen2024} 
we showed that isotropic accretion can become adiabatic when the background is very optically thick, 
or when radiative diffusion timescale is longer than the dynamical timescale. This regime is similar to the well-studied photon trapping limit of black holes \citep{Begelman1978, Thorne1981, Flammang1982, Inayoshi2016, Begelman2017, Wang2021} where accretion rate can be hyper-Eddington.
For clarity, 
in this paper we only focused on the fast diffusion regime when radiation decouples from gas and acts as a reduction in gravity. {The typical radiative diffusion timescale within the accretion flow, which is the accretion lengthscale divided by photon diffusion velocity $\sim c/\tau$, is $ \dfrac{\rho_0 \kappa}{c} \min[ R_{\rm crit, iso}^2, R_H^2] \sim 0.2\Omega^{-1}$ for our $\rho_0$ and electron scattering opacity, so the density or optical depth needs to be at least one order of magnitude larger 
for this timescale 
to be longer than the typical local sound crossing or free fall timescale.}

For adiabatic accretion, 
effect of radiation would be scale free and the short-term evolution outcome 
would be similar to adiabatic CPDs \citep{Fung2019} 
or CPDs with very long cooling timescales \citep{Krapp2022,Krapp2024} (albeit with a different effective adiabatic index), 
where hot gas
form an isentropic pressure supported envelope up to a fraction of the Hill radius, 
while rotation as well as further accretion is suppressed.

However, 
simulations of adiabatic CPD generally neglect the self-gravity of the isentropic envelope. 
In the context of AGN stars, 
\citet{Chen2024} demonstrated that under isotropic conditions, 
this extended adiabatic envelope can undergo gravitational collapse if the stellar companion's entropy, 
is higher than that of the background gas \footnote{which roughly translates to $\Pi_0$ of the disk $\lesssim 0.44$ for a 50 $M_\odot$ star, much smaller than the parameter for our simulations as shown in Table \ref{tab:parameters}.}. 
However, 
the subsequent evolution of this envelope in a disk environment remains unclear. 
When gap-opening effects are strong or when runaway accretion clears out the star's co-orbital region, 
the vertical optical depth can decrease significantly, 
potentially shifting accretion back to a fast-diffusion regime regulated by radiative feedback. 
The exact outcome requires further investigation with long term simulations incorporating self-gravity. 

\section{Summary}
\label{sec:summary}

In this work, 
we performed 3D radiation hydrodynamic simulations of stars embedded in AGN disks, 
in order to determine their accretion structure and rates, under the context where radiation force acts as a reduction in gravity. 
In the isotropic scenario, 
the critical radius 
$R_{\rm crit, iso}$ is determined by radiative feedback from the stellar luminosity as well as the accretion luminosity and $\dot{M} \propto R_{\rm crit, iso}^2$. 
We find that when the disk midplane temperature is high and $R_{\rm crit, iso} < H$ (subthermal), 
accretion remains relatively isotropic, 
and the effective accretion radius aligns with the nominal critical radius. 
When the disk temperature is low, however, $R_{\rm crit} > H$ (superthermal) and strong anisotropy emerges. 
Accretion luminosity tends to escape from the polar region with lowest optical depth driving super-Eddington outflow,  while lower altitudes experience enhanced inflow due to weaker radiative feedback. 
This is in contrast to  
the polar-in, midplane-out vertical flow patterns generally observed 
in circum-planetary disks with strong cooling.
Overall, as the background disk becomes colder and thinner, 
the accretion rate deviates from the isotropic expectation and follows transitions towards a a scaling of $\dot{M} \propto R_{\rm Hill}^2$. 
To facilitate this inflow at low altitudes, which carries significant angular momentum inward, 
viscous stress transports angular momentum 
outward to maintain angular momentum 
conservation, 
which is most likely 
contributed by spiral shocks. 
The measured effective viscosity parameter is on the order of $\alpha_{\rm eff} = 0.1-1$, exceeding or at least comparable to what 
can be sustained by turbulence driven by 
typical magnetorotational or gravitational instabilities.

Our runs span only $2\pi \Omega^{-1}$ on the local scale, 
meaning that for superthermal mass AGN stars, 
gap-opening could further deplete the gas density and affect the long-term accretion rate apart from geometric factors \citep{Li2023}. 
Additionally, 
stellar irradiation may alter the disk's energy balance and temperature, 
effects that our local simulations do not capture. 
A comprehensive model requires global simulations with self-consistent boundary conditions and turbulent heating from MRI or GI. 

Nonetheless, our results support a simple accretion rate scaling (Equation \ref{eqn:ratescaling}),
which long-term 1D stellar evolution calculations can build upon.
Current studies \citep{Cantiello2021, Dittmann2021, AliDib2023, WangJM2023, DittmannCantiello2025, Fryer2025} 
primarily account for accretion rate reduction due to stellar luminosity alone, 
or effectively assuming $R_{\rm crit } \sim (1-\lambda_\star) GM_\star/c_{\rm s, gas}^2$ which can be a poor estimate. 
Additional feedback from accretion luminosity could further reduce the accretion rate, 
and therefore constrain the maximum mass
and evolutionary outcomes of these stars, 
although extra care should be taken in incorporating the forementioned long-term and global effects.
On the other hand, 
based on explorations of \citet{Chen2024}, 
when the surrounding optical depth is large enough and radiation couples with gas,
the system is expected to enter the slow diffusion regime, where adiabatic accretion 
leads to the formation of a pressure-supported isentropic envelope beyond the stellar surface — an effect not considered in any existing literature of long-term modeling of AGN stars. 
Furthermore, 
such an envelope could become self-gravitating and collapse, 
introducing further complexity. 
In a parallel study, we will explore this slow diffusion or adiabatic limit with significantly higher background density.

YXC would like to thank 
Douglas Lin, Wenrui Xu, James Stone and Geoffroy Lesur for helpful discussions. 
We thank Zhaohuan Zhu for sharing his numerical setup for simulating accretion of proto-Jupiter in a protoplanetary disk.
We acknowledge computational resources provided by the high-performance computer center at Princeton University, 
which is jointly supported by the Princeton Institute for Computational Science and Engineering 
(PICSciE) and the Princeton University Office of Information Technology.
Another source of computation time for our simulations is the NSF's ACCESS program (formerly XSEDE) under grants PHY240047 on Purdue's ANVIL supercomputer.



\appendix
\section{Angular-momentum transport by spirals}
\label{sec:appA}

{Since the spirals in our simulations are actually strong shocks, we seek a formula for their angular-momentum flux that is independent of quasilinear approximations. 
We also want this formula to be adaptable to our simulation data, which do not necessarily follow the assumption of strict self-similarity made by \cite{Spruit1987}.
Our approach is heuristic rather than rigorous, 
as we merely want to verify that the spirals we see in the simulation could plausibly account for the observed accretion rate through the disk; 
so we will be satisfied with an order-of-magnitude estimate.}

{Consider first a nonlinear wave in one spatial dimension ($x$) propagating quasi-steadily at constant speed $U$.
We assume that the wave profile is effectively of limited width, and that the distance over which the profile decays and/or $U$ varies significantly is large compared to this width, so that the density $\rho(x,t)$ and fluid velocity $v(x,t)$ depend to a good approximation only on the combination $\xi=x-Ut$.
}

From the continuity equation

\begin{equation}
    \dfrac{\partial \rho}{\partial t} + \dfrac{\partial (\rho v)}{\partial x} \approx \dfrac{d}{d \xi} (-U \rho + \rho v) = 0,
\end{equation}

it follows that {(even if the profile contains shocks)}

\begin{equation}
    v(\xi) = U\left(1-\dfrac{\overline{\rho}}{\rho(\xi)}\right) = U(1-\hat{\rho}^{-1}),
\end{equation}
Where $\hat{\rho}  = \rho/\overline{\rho}$ density {normalized by the mean density of the background, $\bar\rho$, which we treat as a constant.}

{We take the momentum of the wave to be
\begin{equation}\label{eq:waveP}
    P = \int (\rho-\bar\rho)(v-\bar v)\,dx \approx \bar\rho U\int\frac{(\hat\rho -1)^2}{\hat\rho}\,dx.
\end{equation}

Henceforth we measure $U$ in the frame where the background fluid velocity $\bar v=0$.
To complete our 1D model, we identify the momentum flux associated with the integrated momentum \eqref{eq:waveP} as $T^\mathrm{wave}_{xx} = \bar\rho U^2(\hat\rho -1)^2/\hat\rho$.
This reduces to the familiar expression $(c_s\delta\rho)^2/\bar\rho$ in the case of a sound wave or weak shock.}

{To adapt this to a spiral shock in 2D or 3D with pitch angle $\psi$, let $x$ be a local coordinate perpendicular to the shock front, so that the momentum of the wave/shock flows at angle $\psi$ with respect to the local radial direction.
The associated angular-momentum flux is then $rT^\mathrm{wave}_{r\phi}=r\sin\theta T^\mathrm{wave}_{xx}\sin\psi\cos\psi $.
In the approximation that the wave is stationary in the computational frame, the velocity of the shock with respect to the local fluid is $U=-(\bar v_\phi\sin\psi+\bar v_r\cos\psi)$.  We drop the term in $\bar v_r$ because the accretion is slow compared to the orbital speed and $\psi$ turns out not to be especially small.  Here the overbars are interpreted as averages with respect to $\phi$.
Putting this together,
\begin{equation}\label{eq:spiral_amflux}
  \overline{T^\mathrm{wave}_{r\phi}} \approx r\sin\theta\bar\rho\bar v_\phi^2 \sin^3\psi\cos\psi \times
\overline{\frac{(\hat\rho-1)^2}{\hat\rho}}\,.
\end{equation}
Multiplication of this last expression by $2\pi r^2\sin\theta$ and integration over colatitude $\theta$ estimates $\dot J_\mathrm{wave}(r)$, the rate at which the spirals carry angular momentum through the sphere of radius $r$.
}

{Although wave transport is not entirely equivalent to a viscosity, we compare eq.~\eqref{eq:spiral_amflux} to the $\alpha$ prescription for a viscous keplerian disk, $T_{r\phi}=\tfrac{3}{2}\alpha r\bar\rho c_s^2$ (at the midplane $\theta=\pi/2$).  We obtain}
\begin{equation}
    \alpha_{\rm wave} \approx \dfrac23\left(\dfrac{\overline{v}_\phi}{c_s}\right)^2\sin^3\psi \cos\psi \overline{\left[\dfrac{(\hat{\rho} - 1)^2}{\hat{\rho}}\right]}
    \label{eqn:alphawave}
\end{equation}

As an example, we apply this framework to the time-average midplane density profile of run \texttt{T3e4}, ignoring the $\theta$ dimension since for this superthermal case most materials are at low altitude. 
We first try to fit a logarithmic spiral $\hat{\rho} (\phi,r) = f(\phi + \cot\psi \ln r)$ to the density perturbation by minimizing the objective function

\begin{equation}
    E(\gamma) = \sum_{j} \sum_{i} \left[ \cot\psi \times \frac{\hat{\rho}(\phi_i, r_j) - \hat{\rho}(\phi_{i+1}, r_j)}{\phi_{i+1} - \phi_i} - \frac{\hat{\rho}(\phi_i, r_{j+1}) - \hat{\rho}(\phi_i, r_j)}{\ln r_{j+1} - \ln r_j} \right]^2
\end{equation}

over the accretion flow region beyond the stellar convective zone $r> 100R_\odot$. 

\begin{figure}
    \centering
    \includegraphics[width=0.48\textwidth]{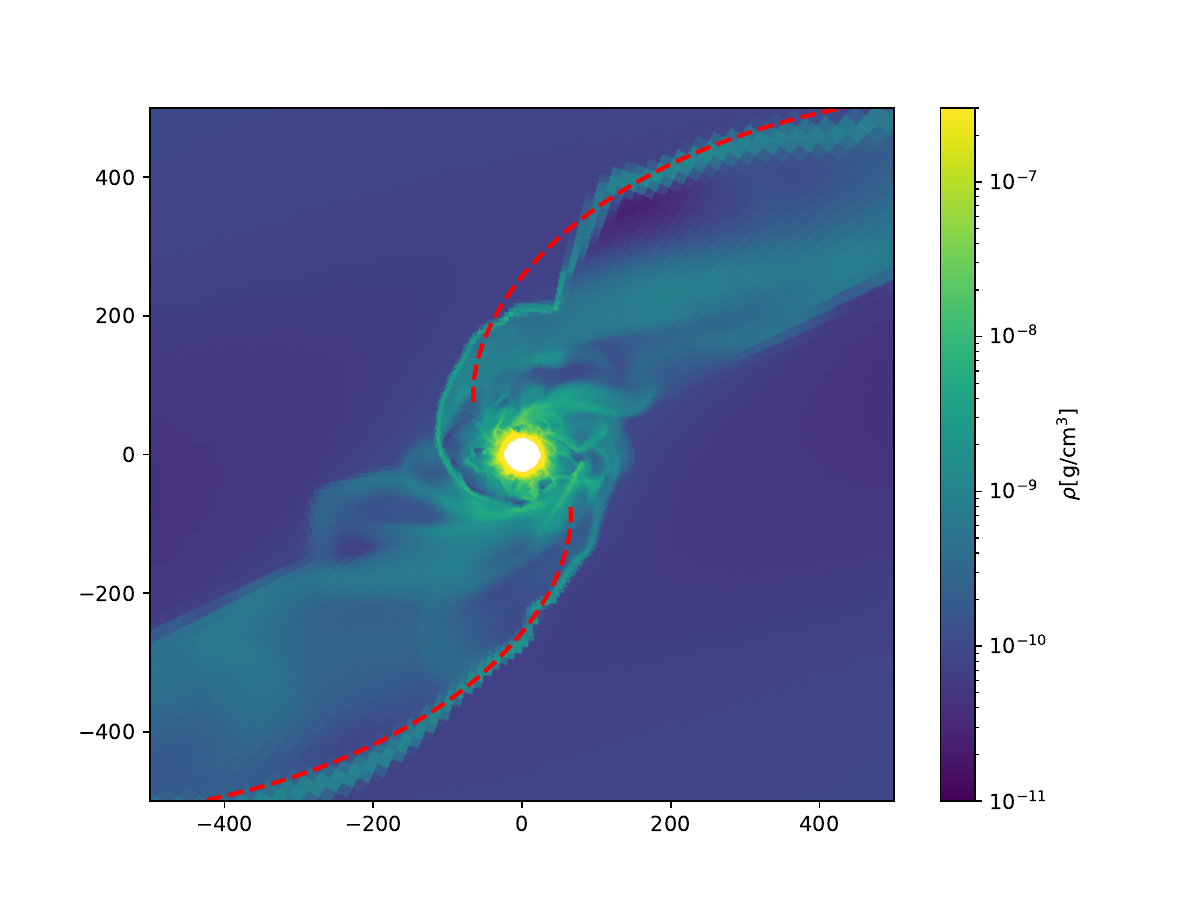}
        \includegraphics[width=0.48\textwidth]{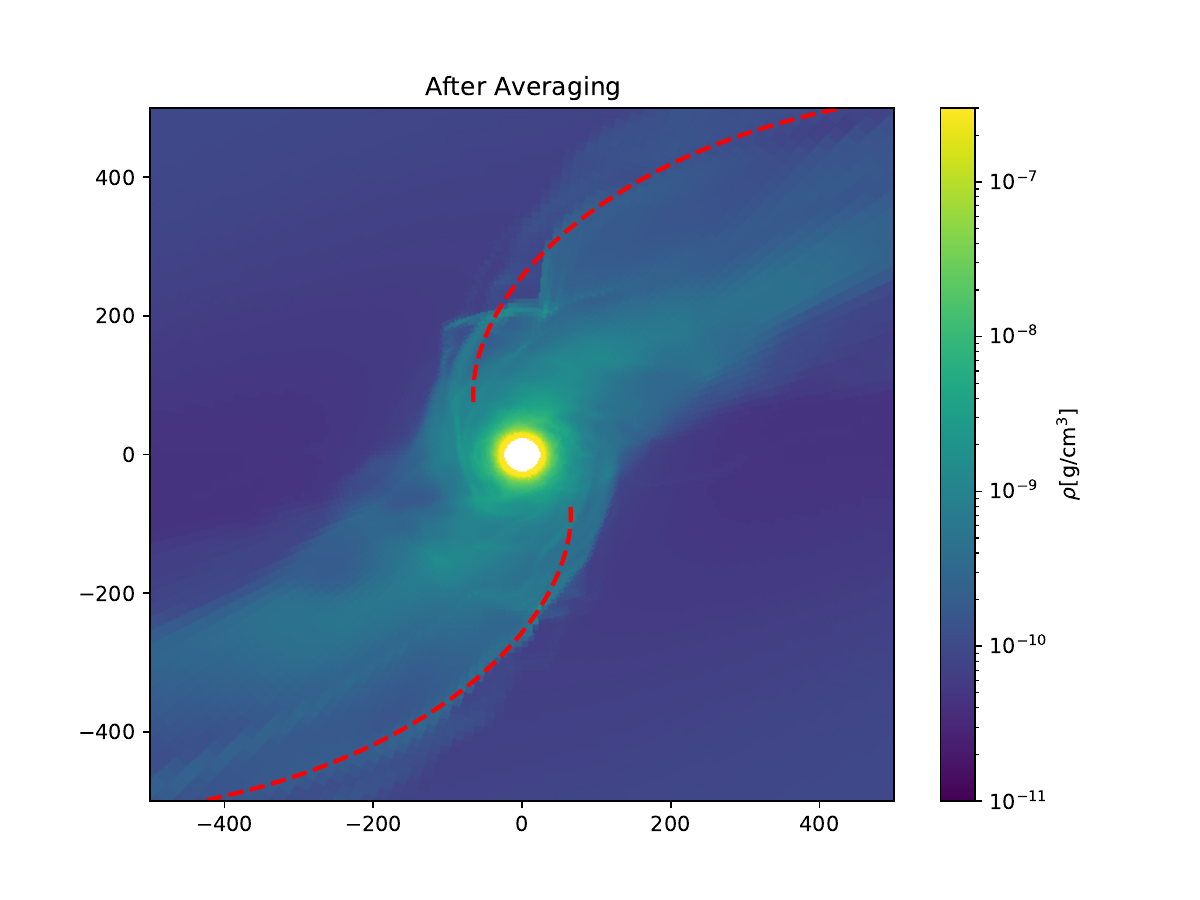}
    \caption{Midplane density distribution in a snapshot and in the data averaged over a dynamical timescale $\Omega^{-1}$ for run \texttt{T3e4}. 
    Red dashed lines: fits of spiral wave front using $\psi = 53^\circ$. }
\label{fig:spiralfit}
\end{figure}

We sketch out spirals defined by the best-fit pitch angle $\psi = 53^\circ$ and 
find that 
we can accurately depicts the wave structure for midplane density distribution of individual snapshots, see left panel of Figure \ref{fig:spiralfit}. 
Interestingly, when examining the time-averaged data (right panel of Figure \ref{fig:spiralfit}), 
the spirals
does not perfectly align with the location of largest density contrast. 
This discrepancy probably arises from statistical smearing caused by turbulence 
in the time-averaged data.

\begin{figure}
    \centering
    \includegraphics[width=0.5\textwidth]{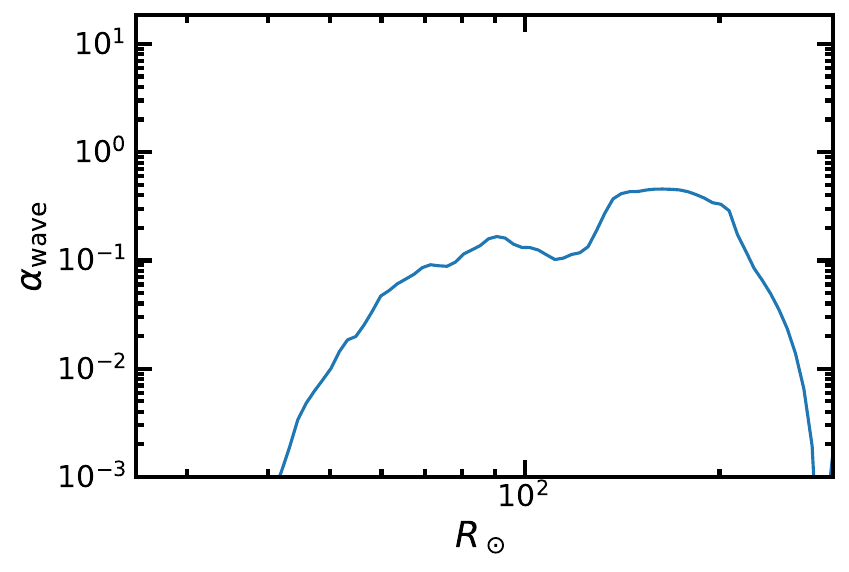}
    \caption{Radial profile of midplane effective viscosity parameter calculated from Equation \ref{eqn:alphawave} for the run \texttt{T3e4}.}
\label{fig:alphawave}
\end{figure}

By substituting the value of $\psi$ into 
Equation \ref{eqn:alphawave}, 
we calculate the radial profile of effective $\alpha_{\rm eff}$ profile. 
The result is plotted in Figure \ref{fig:alphawave}, 
indicating that for the measured density and azimuthal velocity profiles, 
this spiral arm can indeed generate up to an effective viscosity parameter of $\alpha_{\rm wave} \sim 0.1-1$. 
Despite the simplifying approximations in our analysis, 
we believe this offers sufficient evidence that spiral arms can contribute effectively to angular momentum transport.

\bibliography{sample631}{}
\bibliographystyle{aasjournal}



\end{CJK*}
\end{document}